\documentclass[twocolumn,amsmath,amssymb,prl,superscriptaddress]{revtex4}

\usepackage{amsmath}
\usepackage{graphicx}
\usepackage{color}
\usepackage[normalem]{ulem}
\usepackage{natmove}
\usepackage{float}
\usepackage[colorlinks=true, allcolors=blue]{hyperref}

\begin{document}
	
\title{Machine learning determination of atomic dynamics at grain boundaries}

\affiliation{Department of Physics and Astronomy, University of Pennsylvania, Philadelphia, PA}
\affiliation{Department of Materials Science and Engineering, University of Pennsylvania, Philadelphia, PA}
\affiliation{Department of Mechanical Engineering and Applied Mechanics, University of Pennsylvania, Philadelphia, PA}

\author{Tristan A. Sharp}
\affiliation{Department of Physics and Astronomy, University of Pennsylvania, Philadelphia, PA}
\author{Spencer L. Thomas}
\affiliation{Department of Materials Science and Engineering, University of Pennsylvania, Philadelphia, PA}
\author{Ekin D. Cubuk}
\affiliation{Department of Materials Science, Stanford University, Stanford, CA}
\author{Samuel S. Schoenholz}
\affiliation{Google Incorporated, Mountain View, CA}
\author{David J. Srolovitz}
\affiliation{Department of Materials Science and Engineering, University of Pennsylvania, Philadelphia, PA}
\affiliation{Department of Mechanical Engineering and Applied Mechanics, University of Pennsylvania, Philadelphia, PA}
\author{Andrea J. Liu}
\affiliation{Department of Physics and Astronomy, University of Pennsylvania, Philadelphia, PA}

\begin{abstract}

In polycrystalline materials, grain boundaries are sites of enhanced atomic motion, but the complexity of the atomic structures within a grain boundary network makes it difficult to link the structure and atomic dynamics.
Here we use a machine learning technique to establish a connection between local structure and dynamics of these materials. 
Following previous work on bulk glassy materials, we define a purely structural quantity, softness, that captures the propensity of an atom to rearrange.  This approach correctly identifies crystalline regions, stacking faults, and twin boundaries as having low likelihood of atomic rearrangements, while finding a large variability within high-energy grain boundaries. As has been found in glasses 
{\normalfont [Schoenholz SS, Cubuk ED, Sussman DM, Kaxiras E, and Liu AJ. (2016) \emph{Nat Phys}, 12(5):469-471; 
Schoenholz SS, Cubuk ED, Kaxiras E, and Liu AJ. (2017) \emph{Proc Natl Academ Sci USA}, 114(2):263-267; 
Sussman DM, Schoenholz SS, Cubuk ED, Liu AJ. (2017) \emph{Proc Natl Academ Sci USA}, 114(40):10601-10605]
}, 
the probability that atoms of a given softness will rearrange is nearly Arrhenius. This indicates a well-defined energy barrier as well as a well-defined prefactor for the Arrhenius form for atoms of a given softness. The decrease in the prefactor for low-softness atoms indicates that variations in entropy exhibit a dominant influence on the atomic dynamics in grain boundaries. 
\end{abstract}

\maketitle

Atomic rearrangements, in which atoms overcome energy barriers to change neighbors, underpin the dynamics of grain boundaries (GBs) in polycrystalline materials that are ultimately responsible for such phenomena as grain growth (GB migration), GB diffusion, GB sliding, emission/absorption of lattice dislocations, and point defect sink behavior. 
Assessing the atomic-scale dynamics of GBs from their structure in realistic polycrystalline materials is inherently complicated. This complexity is associated with both the highly degenerate nature of GB structures~\cite{sutton1996interfaces,vitek1983multiplicity,Han2016Grain} as well as the interconnected nature of the GB network within polycrystalline microstructures (e.g., associated with GB junctions, compatibility, \emph{etc}). The ensemble of possible configurations produces a nearly continuous spectrum of energies and a correspondingly large set of local structures~\cite{Han2016Grain}.
The atomic structure at the GB exhibits some order imparted by the contiguous grains, but also becomes trapped in metastable minima in a complex energy landscape reminiscent of that of a glass~\cite{zhang2009grain,HOU2015177,Han2016Grain}.
This suggests that methods appropriate for glass dynamics may be fruitful for understanding the dynamics internal to GBs in polycrystals.
 
Previously, it has been shown that low-frequency quasilocalized vibrational modes can be used to identify atoms that are likely to rearrange in both glasses~\cite{manning2011vibrational,schoenholz2014understanding} and GBs~\cite{rottlerPRE2014}.
Here we push the analogy between GBs and glasses further by using a machine learning analysis, originally developed to study atomic rearrangements in glasses, to characterize the local atomic rearrangements in molecular dynamics (MD) simulations of polycrystalline solids.
The rearrangements occur as atoms thermally fluctuate over local barriers between metastable sites.
On long time scales, asymmetries in these dynamics give rise to kinetic phenomena such as GB migration, creep, and defect emission, but here we consider the individual atomic rearrangements within the GB.

Because defect microstructures are commonly spatially extended, it is not clear that atomic rearrangements can be characterized in terms of only local structural information at the atomic scale. Here we demonstrate that structural information within a few atomic diameters is indeed sufficient to predict these atomic-scale rearrangements and characterize particles in terms of a single continuous scalar variable, called ``softness," which captures the relevant properties of the local atomic environment. Remarkably, we find that particles of a given softness are characterized by a well-defined common energy barrier to rearrangements in polycrystals, just as was previously discovered for glassy liquids~\cite{Schoenholz2016Structural}. Thus, our results translate into a spatial map of the energy barriers to rearrangements. Our findings suggest that it is possible to characterize much GB dynamical behavior using only local, atomic-scale structural information.

\begin{figure*}
	\begin{center}
		\includegraphics[width=0.215\linewidth]{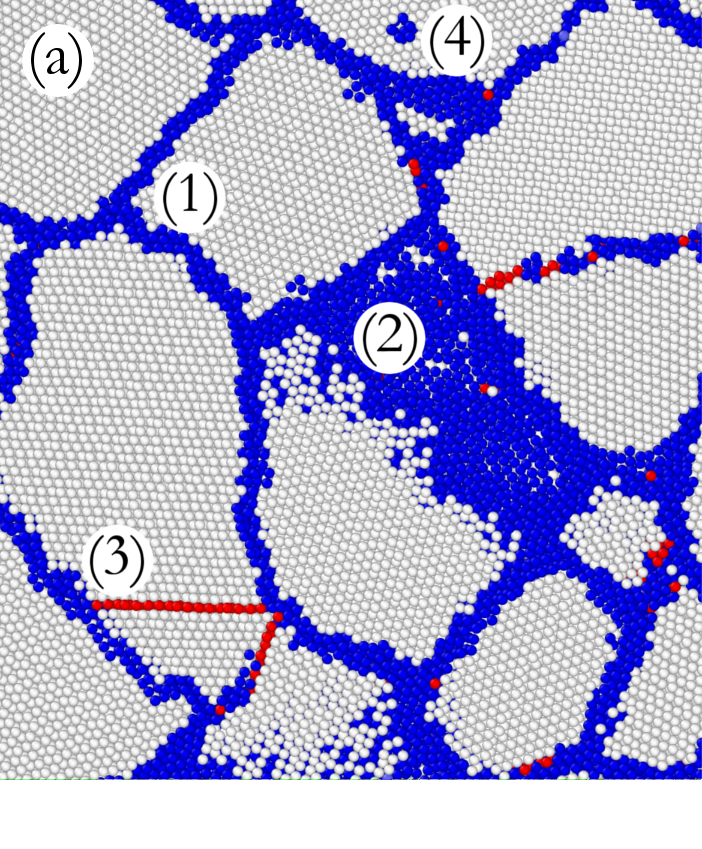}
		\includegraphics[width=0.215\linewidth]{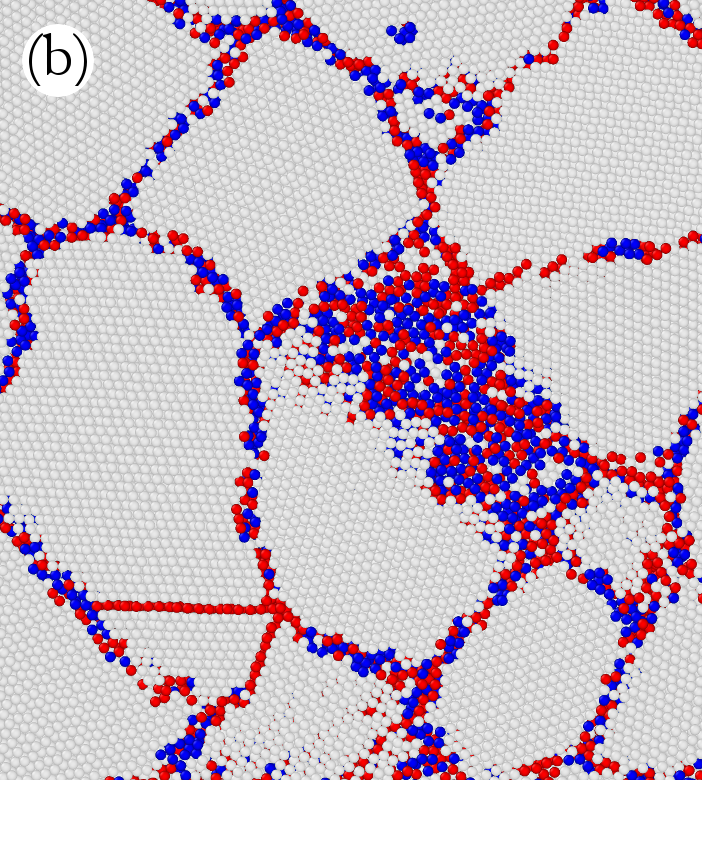}
		\includegraphics[width=0.215\linewidth]{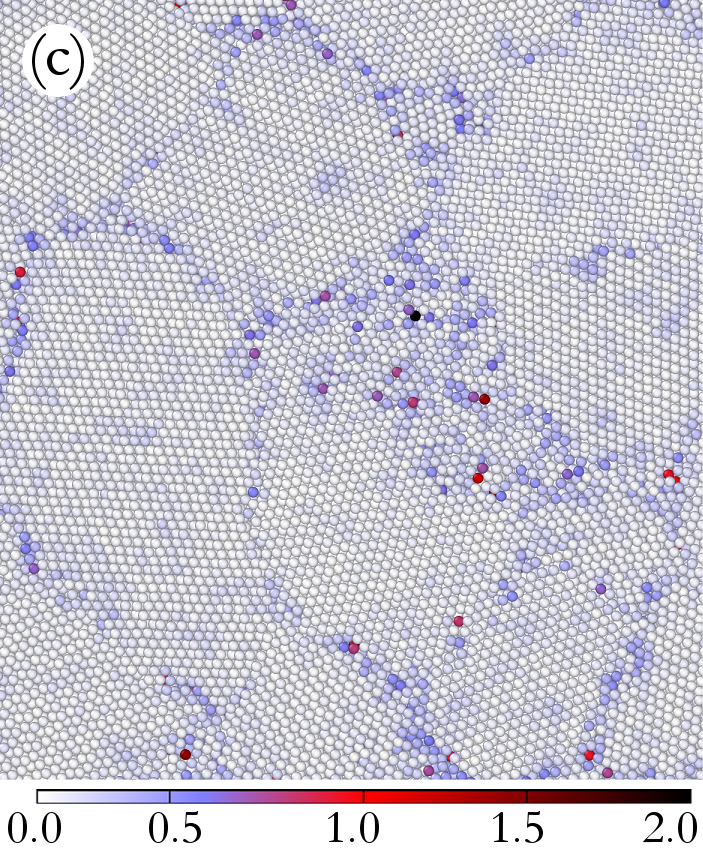}
		\includegraphics[width=0.215\linewidth]{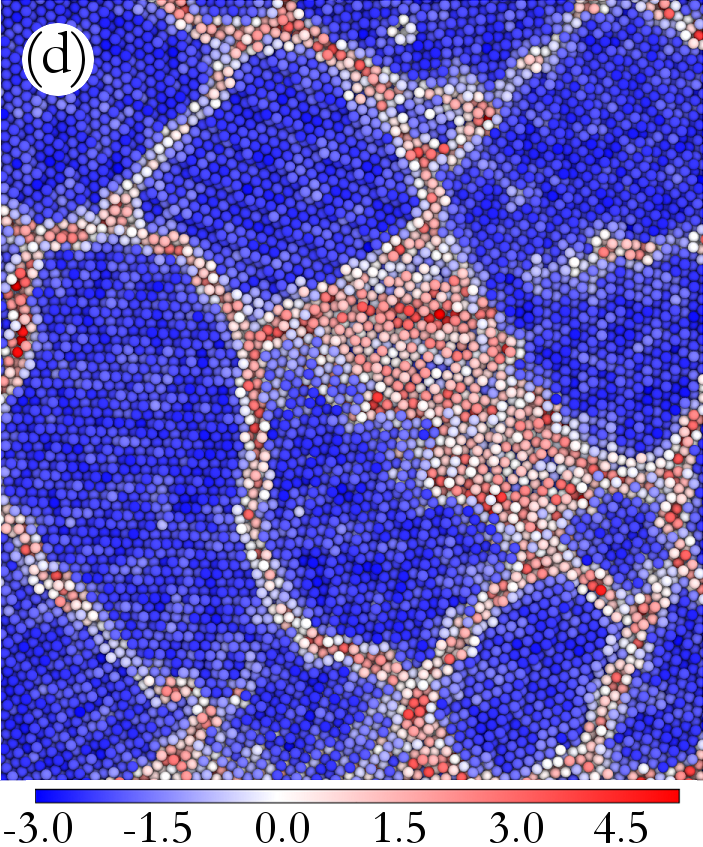}
		\caption[]{\label{fig1}
			A small region of a cross-section through the three-dimensional polycrystalline aluminum microstructure at $T=463$ K as visualized by (a) common neighbor analysis and (b) a Voronoi topology analysis~\cite{Lazar2015}. Grain interiors are locally FCC (white atoms), while stacking faults and twin boundaries are visible as locally hexagonal close-packed (HCP) structures (red atoms), and atoms within general GBs are neither FCC nor HCP  (blue atoms).  Positions are averaged over a $0.1$~ps window. 
			(c) Same atoms colored by their instantaneous value of $p_{hop}$ ($\text{\AA}^2$) as indicated by the color bar.
			(d) Same atoms colored according to their softness, as indicated by the color bar.
		}
	\end{center}
\end{figure*}

\section*{Methods}

\begin{figure}
	\begin{center}
		\includegraphics[width=0.95\linewidth]
		{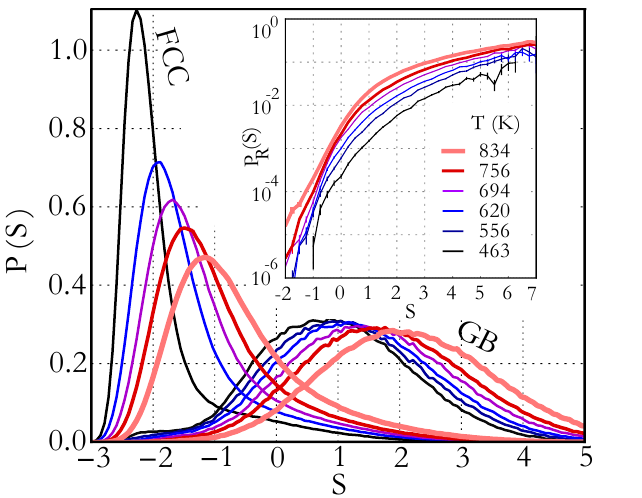}
		\caption[]{
			\label{figpdfsoft}
			\label{PR_of_S}
			The distribution of softness, $S$, within grain interiors and at GBs.
			The interiors are distinguished from GBs on the basis of their Voronoi topologies. About 89\% of atoms are in an FCC environment.
			(Inset)
			The probability to rearrange, $P_R$, rises monotonically with softness, $S$, over several orders of magnitude for all temperatures.
		}
	\end{center}
\end{figure}

Large-scale molecular dynamics simulations~\cite{PlimptonFast1995} of nanocrystalline aluminum are performed as in Ref.~\citenum{Thomas2016}, producing a network of interacting GBs. The geometry is initialized using a random (Poisson point process) Voronoi tesselation and curvature-flow grain growth algorithm~\cite{Lazar2011,Lazar2015}. Atomic interactions model aluminum using the embedded atom method~\cite{MendelevAluminumEAM2008}. The grains are face-centered cubic (FCC) with nominal nearest-neighbor distance $d$ of approximately $2.8$~\AA, and a melting temperature of $T_m = 926$ K.  The system of 4 million atoms is simulated at a fixed temperature, $T$, using a Nos\'e-Hoover thermostat.  The atomic positions are averaged over 0.1 ps intervals to reduce the highest frequency atomic vibrations. All data and code are archived and available upon request.

The microstructure of the simulated nanocrystalline aluminum can be visualized using an adaptive common neighbor analysis \cite{Stukowski2010Ovito} (CNA, Fig.~1 (a)) or Voronoi topology analysis~\cite{Lazar2015,lazar2015topological} (VoroTop, Fig.~1 (b)) to distinguish grains from GBs.
Both CNA and VoroTop assess local crystalline structure around an atom using the relative positions of the atom's neighbors.
CNA defines neighbors as those atoms within a cutoff distance somewhat larger than $d$ and compares which neighbors are also neighbors of each other.  Lists of shared neighbors are then compared to those generated from ideal crystalline lattices.
In contrast, VoroTop does not impose a cut-off distance, but analyzes the atomic Voronoi cell structure to determine similarity to cells from perturbed crystalline lattices.
This method more readily identifies crystalline environments, recognizing 89\% of atoms in the simulation in Fig.~1 as FCC, compared to 78\% by CNA.
The VoroTop analysis is also much more robust against thermal fluctuations, so that the identified GB regions do not grow significantly with temperature, up to $0.8$ $T_m$.

Our aim is to characterize the propensity for an atom to rearrange.
We follow a procedure designed for glassy systems~\cite{cubuk2015identifying,Schoenholz2016Structural,CubukSchoenholz2016,schoenholz2017relationship}
that uses a support vector machine (SVM)~\cite{Chang2011SVM} (a supervised machine learning technique) to identify correlations between the local structure around each atom and its rearrangements. 
Complete details are given in the SI Appendix.
A descriptor, or fingerprint, of the local structure around each atom is calculated by evaluating a set of 
so-called structure functions~\cite{PhysRevB.87.184115}. Structure functions are functions of
the relative positions of the atoms out to a cutoff distance $d_0 = 5.0 \text{ \AA}$ from a central atom, 
about twice the interatomic spacing. 
We use two types of structure functions to ensure that different atomic environments lead to different fingerprints~\cite{BehlerParinello2007,PhysRevB.87.184115}.
The first type depends on the radial distribution of neighbors from the central atom, whereas the second type encodes the distribution of triangles of varying angles formed by the central atom with pairs of neighbors. (The complete  functions are given in the SI Appendix.)

Next, we identify which atoms rearrange in the course of the dynamical simulation. For this, we use a standard atom-based quantity, $p_{hop}(t)$ ~\cite{Keys2011Excitations,SmessaertRottler2013}; $p_{hop}(t)$ becomes large when the atom moves a large distance compared with the atomic vibration amplitude (see SI Appendix). Examination of Fig.~1(c) shows that  $p_{hop}$ is only large at the GBs (marked as red or blue atoms in Figs.~1(a) and (b)).
Rearrangements, are defined as events that exceed the threshold $p_{hop} > p_c =1.0 \text{ \AA}^2$.
There is also considerable variation within and between GBs; although 22\% of the atoms are in disordered environments in our microstructure according to CNA, at any instant fewer than 0.06\% of atoms are rearranging and only 0.5\% of the atoms have $p_{hop}>0.5 \text{ \AA}^2$.

To correlate rearrangements to atomic environments, we construct a training set for the supervised machine learning algorithm.
We choose atoms to put into the training set based on whether they arrange within a 200 fs window (we label these $y_i=1$) or do not rearrange over a much longer time (1.8 ps) period (we label these $y_i=-1$). 

The local structural environment around any atom is described as a point in a high-dimensional space in which each orthogonal axis corresponds to possible values of a different structure function. The SVM identifies the hyperplane in this space that best separates the two groups of atoms in the training set (generalized linear regression). We find that 96\% of atoms with $y_i=1$ fall on what we define as the ``positive" side of the hyperplane, and 90\% of the  atoms with $y_i=-1$ fall on the other (``negative'') side of the hyperplane. This justifies using a hyperplane rather than a more general surface to separate the two groups.

\section*{Results}

\subsection*{Linking structure and dynamics}
For any atom, the point characterizing its local structural environment can then be compared to the hyperplane in structure-function space. The signed normal distance from the point to the hyperplane defines the value of the ``softness,'' $S$~\cite{Schoenholz2016Structural}.
Figure~1(d) shows the system with each atom colored according to its softness. 
Atoms with large positive softness (red) tend to lie in GBs, while those with large negative softness (blue) lie within the grain interiors, as expected.
In the SI Appendix we find that softness is partially correlated with other structural quantities such as free atomic volume, but that softness is considerably more predictive of whether atoms rearrange.

We note several interesting features in Fig.~1 associated with the considerable variation in softness along the GBs in the system.
Comparing Fig.~1(a) with 1(d), we see that the GB above the label (1) has small softness relative to the other GBs in the system.
Examination of the structure of that GB demonstrates that it is a large-angle twist GB lying along a close-packed \{111\} plane. 
Such a GB provides only very localized  distortions to the crystal structure of the grains, and few atomic rearrangements occur there. 

At label (2), the viewing plane is nearly co-planar with the GB, showing both structural and dynamic heterogeneity within the GB.
The position in the microstructure labeled (3) shows the intersection of a coherent twin boundary with a more general GB.
The twin boundary is no softer than the grain interior.
Its intersection with the GB, however, changes the GB character (misorientation) and the resulting softness such that the segment of the GB above the intersection with the twin boundary is significantly softer than the segment of the GB just below the intersection.
A lattice vacancy is seen near label (4) in Fig. 1; it is softer than the surrounding lattice but not as soft as many of the GB sites.
This is consistent with earlier work that analyzed vibrational modes and found that vacancies were more mechanically stable than GBs~\cite{ChenPRE2013}.
Indeed, we find vacancies have softness $S \approx 0$. 

Figure~\ref{figpdfsoft} puts these results into context. 
Here we decompose the distribution of softness from the polycrystalline system into contributions from the grain interiors (i.e., FCC regions as determined by the Voronoi topology analysis) and all other regions. The least soft atoms are FCC while the softest are associated with defects.
Softness is more homogeneous in the grain interiors than in the regions that VoroTop defines as disordered (mostly GBs).
The softest atoms tend to lie near the center of the GBs.
In both regions, the distribution shifts slightly with temperature, reflecting a shift in the distribution of local atomic configurations due to thermal distortions of the crystalline grains as well as an increase in the effective GB thickness.

Although we used only a binary classification to define the hyperplane, the magnitude of the softness is predictive of dynamics. 
We establish this by studying the probability that an atom rearranges, $P_R(S)$, defined as the time average of the fraction of atoms in the system with a given $S$ with $p_{hop} > p_c$.
Figure~\ref{PR_of_S} (inset) shows that the fraction of atoms with a given $S$ which will rearrange, $P_R(S)$, is a strongly increasing function of softness $S$. $P_R(S)$ increases monotonically with $S$, and $S=1$ atoms are more than 100 times more likely to rearrange than $S=-1$ atoms. At large $S$, $P_R(S)$ saturates as the fraction of hopping atoms cannot exceed $1$.

\subsection*{Extracting the energy barriers of atomic rearrangements}

\begin{figure}
	\begin{center}
		\includegraphics[width=1\linewidth]{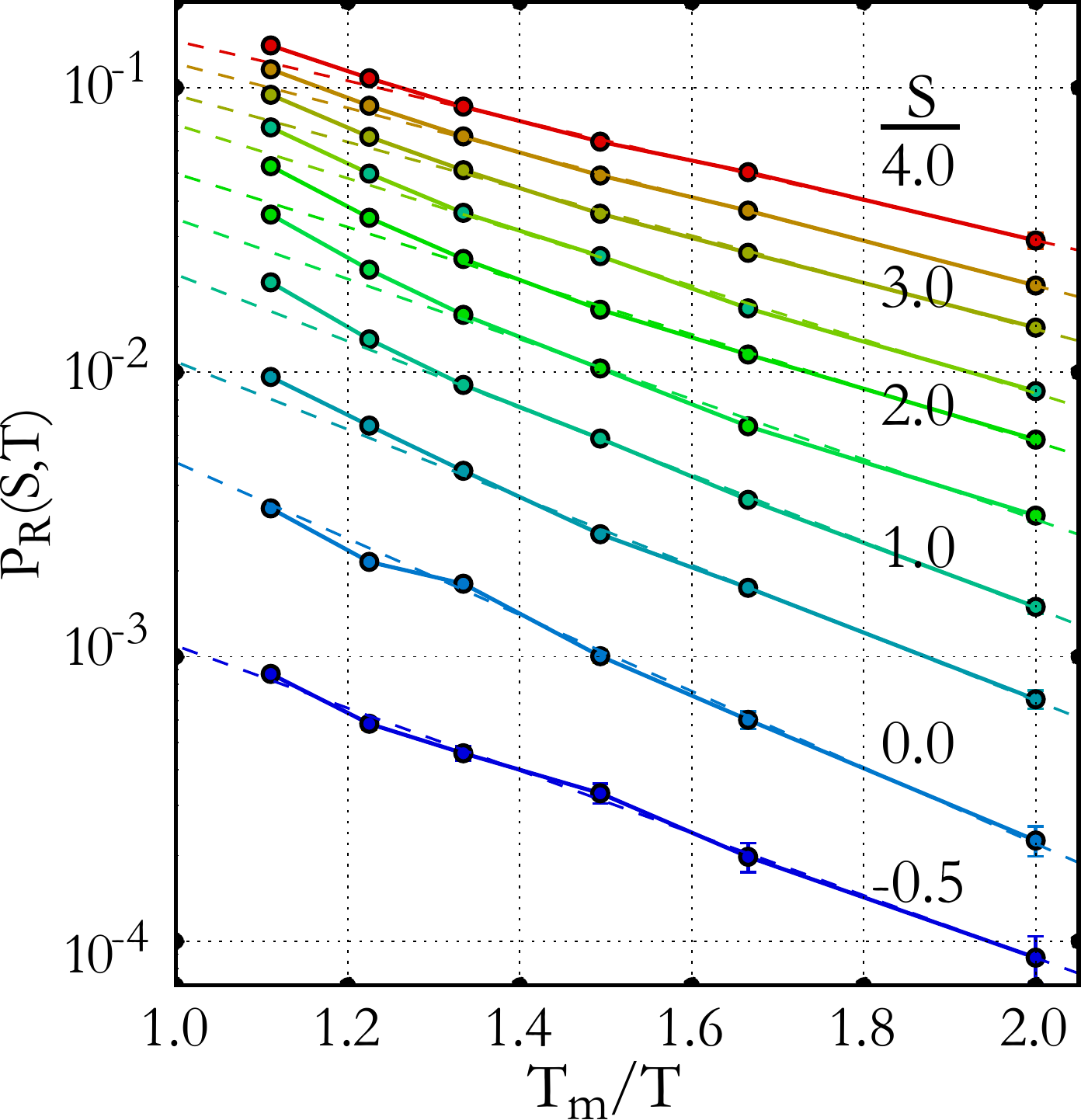}	
		\caption[]{\label{figarrhenius}
			The probability of rearrangement exhibits Arrhenius behavior at temperatures below 0.80 $T_m$.
			Fits to the lower four temperatures are shown as dashed lines.}
	\end{center}
\end{figure}

In earlier work on glassy systems, it was discovered that the probability that an atom rearranges, $P_R(S)$, is Arrhenius for each value of $S$~\cite{Schoenholz2016Structural,schoenholz2017relationship,Sussman2017surface}, implying a well-defined energy barrier associated with rearrangements, $\Delta E(S)$.
We therefore study the temperature dependence of $P_R(S)$.
Figure~\ref{figarrhenius} shows that the temperature dependence of $P_R(S)$ is indeed well-described as Arrhenius at temperatures $T < 0.8$ $T_{m}$,
showing that softness 
reflects the energy barrier for an atom to rearrange.

\begin{figure}
	\begin{center}
		\includegraphics[width=1\linewidth]{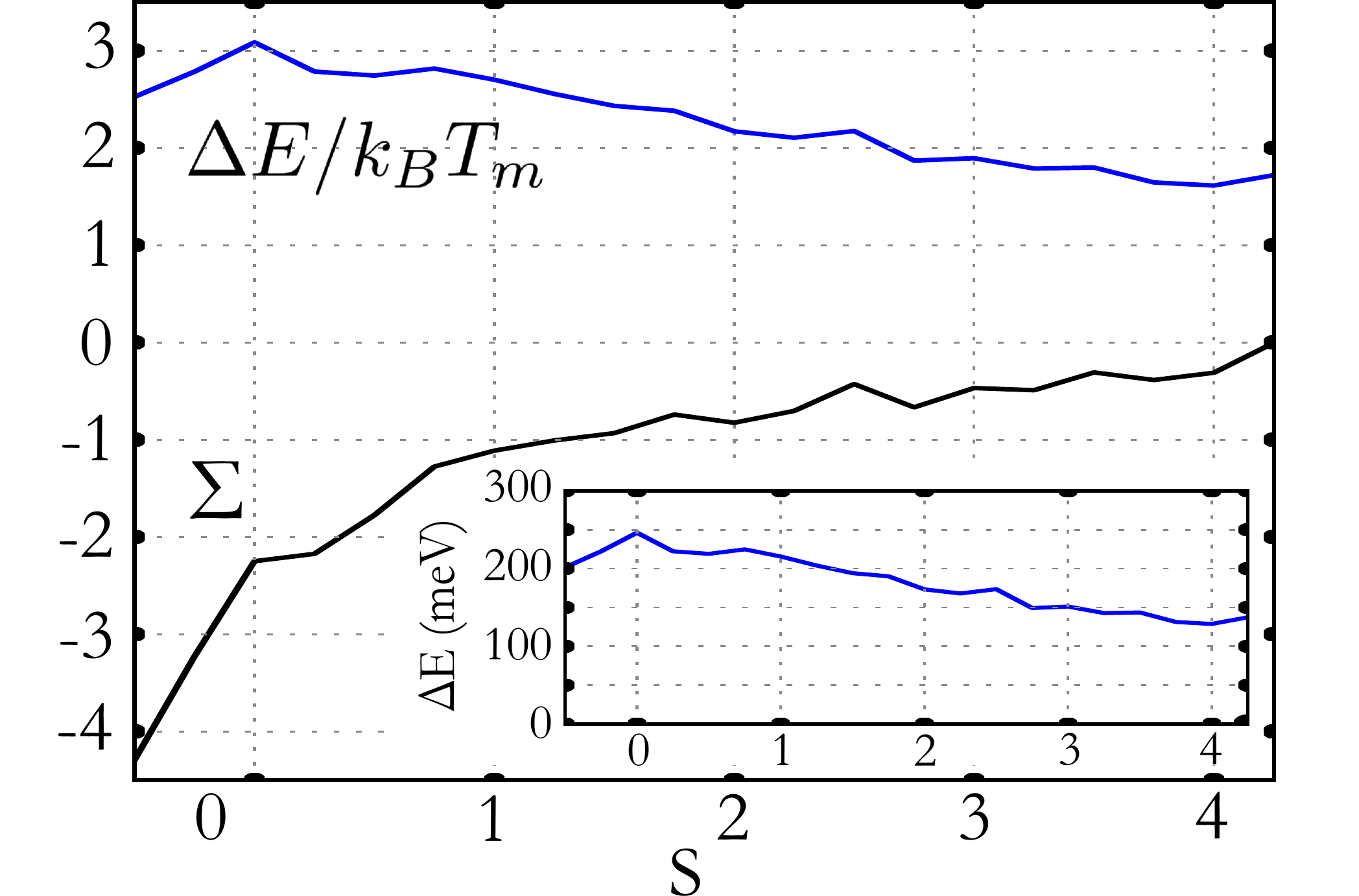}
		\caption[]{\label{figSigmaEbarrier}
			The extracted energy barrier scaled by the thermal energy at the melting temperature, $\Delta E/k_B T_m$ (blue), and $\Sigma$ (black) extracted from the rearrangement probability (Eq.~\ref{eq:arrhenius}). Inset: $\Delta E$ vs.~$S$.  The energy barrier varies slightly with $S$ while $\Sigma$ increases strongly with $S$.
		}
		
	\end{center}
\end{figure}

The Arrhenius form implies that the probability to rearrange can be written
\begin{equation}
\label{eq:arrhenius}
P_R(S) = \text{e}^{\Sigma(S)}\text{e}^{-\Delta E(S)/k_B T}  .
\end{equation}
Figure~\ref{figSigmaEbarrier} indicates that the effective energy barriers decrease slightly with softness, changing by less than a factor of two over the observed range of softness.
The magnitude of the energy barrier is consistent with nudged-elastic band calculations of structural transitions between specific metastable states in  bicrystals using the same interatomic potential employed here~\cite{burbery2016thermo}.
The dominant energy barriers, of order 100 meV, are large compared with applied elastic stress (the yield stress of Al is $\sim 10$ MPa or 1 meV). This shows that these atomic rearrangements will be largely thermally activated, applied stresses have only little effect, and suggests that asymmetries in transition rates underlie stress-driven microstructure evolution. 

Changing softness has much more of an effect on the prefactor in the Arrhenius relationship, $\Sigma$, than it does on the barrier $\Delta E$ 
(see Eq.~\ref{eq:arrhenius} and Fig.~\ref{figarrhenius}).
For a thermally-activated process, $\Sigma$ can be viewed as a generalized attempt frequency, or alternatively as an entropic contribution to the free energy barrier. The increase of $\Sigma$ with softness suggests that soft atoms have more directions for rearrangement or more paths that can take them to the transition state.
This is in contrast to bulk glasses which found that both $\Delta E$ and $\Sigma$ decreased with increasing $S$.
We therefore see that the reason that softer atoms are more likely to rearrange in grain boundaries is that they are the ones with slightly lower energy barriers \emph{and} substantially increased $\Sigma$.

We consider the origin of this difference from bulk glasses. In polycrystals, and unlike in bulk glasses, there are crystalline grains that restrict the rearrangements.
Atoms in the crystalline grain mostly cannot participate in rearrangements due to the large potential energy barrier to move them significantly. With fewer atoms able to participate, the number of possible rearrangement trajectories (and $\Sigma$) is reduced.  The SI Appendix shows explicitly that low softness ($S \approx 0$) atoms are also in the most crystal-like neighborhoods, as indicated by Voronoi volume, radial symmetry, and potential energy.
These crystalline neighborhoods evidently restrict $\Sigma$ sufficiently to have a large impact on the GB dynamics, leading to low-$\Sigma$, low-$S$ atoms.

The best fit lines in Fig.~\ref{figarrhenius} indicate no common intersection point. This implies that there is no one temperature at which rearrangement dynamics become independent of softness, or local structure. This contrasts with the behavior of glassy systems, which display a common intersection point~\cite{Schoenholz2016Structural} at what is known as the onset temperature. This temperature marks the onset of features associated with supercooled liquids, such as non-exponential relaxation, a non-Arrhenius dependence of the relaxation time on temperature and kinetic heterogeneities. The fact that there is no indication of an onset temperature for this system indicates that GBs, despite exhibiting some glass-like properties~\cite{zhang2009grain} also exhibit behavior that is very unlike glasses.

\subsection*{Analysis of SVM}

We next analyze which structure functions (SFs) are most responsible for the softness of the atom. 
Recall that the structural quantities we use to characterize the local environment of each atom fall into two classes; radial SFs depend only the radial distribution of atoms from the central atom, and angular SFs additionally depend on the angles formed with pairs of neighboring atoms around the central atom.
In bulk glasses, angular SFs were unimportant compared to the radial SFs, and in fact softness was almost entirely attributable to the number of neighboring atoms at the distances of the first peak and valley of the pair correlation function of the material, $g(r)$\cite{Schoenholz2016Structural}. In grain boundaries, one may expect increased importance of angular SFs since they are sensitive to the relative orientations of the lattices, \emph{ie} the crystallography of the grain boundary.

To connect with studies on bulk glasses, we exclude crystalline atoms from the training set using CNA, although we find that this does not significantly change the relative importance of most structure functions. Excluding crystalline atoms does however decrease the fraction of training set atoms which are accurately classified from 93\% to 79\%, since now the SVM discriminates solely between disordered atoms.

We use recursive feature elimination (RFE) to determine the important features.  In RFE, the SVM is first trained using all SFs.  Then the SF that contributes least weight to the hyper-plane is identified and eliminated.  Training is repeated, iteratively identifying and eliminating the least important SF, simplifying the fingerprint while attempting to retain the highest accuracy.

Figure~5 shows the decrease in accuracy as the number $M$ of SFs decreases. The color of the symbol indicates the class of the least-significant SF which then becomes eliminated.  All radial functions are eliminated quickly (red hollow circles), decreasing accuracy $f_a$ only slightly to 77\%. Of 72 total features, the last 47 features to be eliminated (and therefore, the top 47 most important quantities) are all angular functions. This is in stark contrast to bulk glasses, where all of the angular functions could be eliminated entirely with less than a 2\% cost in accuracy\cite{Schoenholz2016Structural}. 
14 angular functions are sufficient to retain 77\% accuracy, and accuracy remains near 69\% using the sole SF identified as most important.
The increased importance of angular information here may seem natural given the role of the degree and relative orientation of the nearby crystallinity in allowing rearrangements.
The contribution of each specific SF to softness is quantified in the SI Appendix.
The properties of the specific structures that lead to rearrangements should be explored further.

\begin{figure}
	\begin{center}
		\includegraphics[width=1\linewidth]{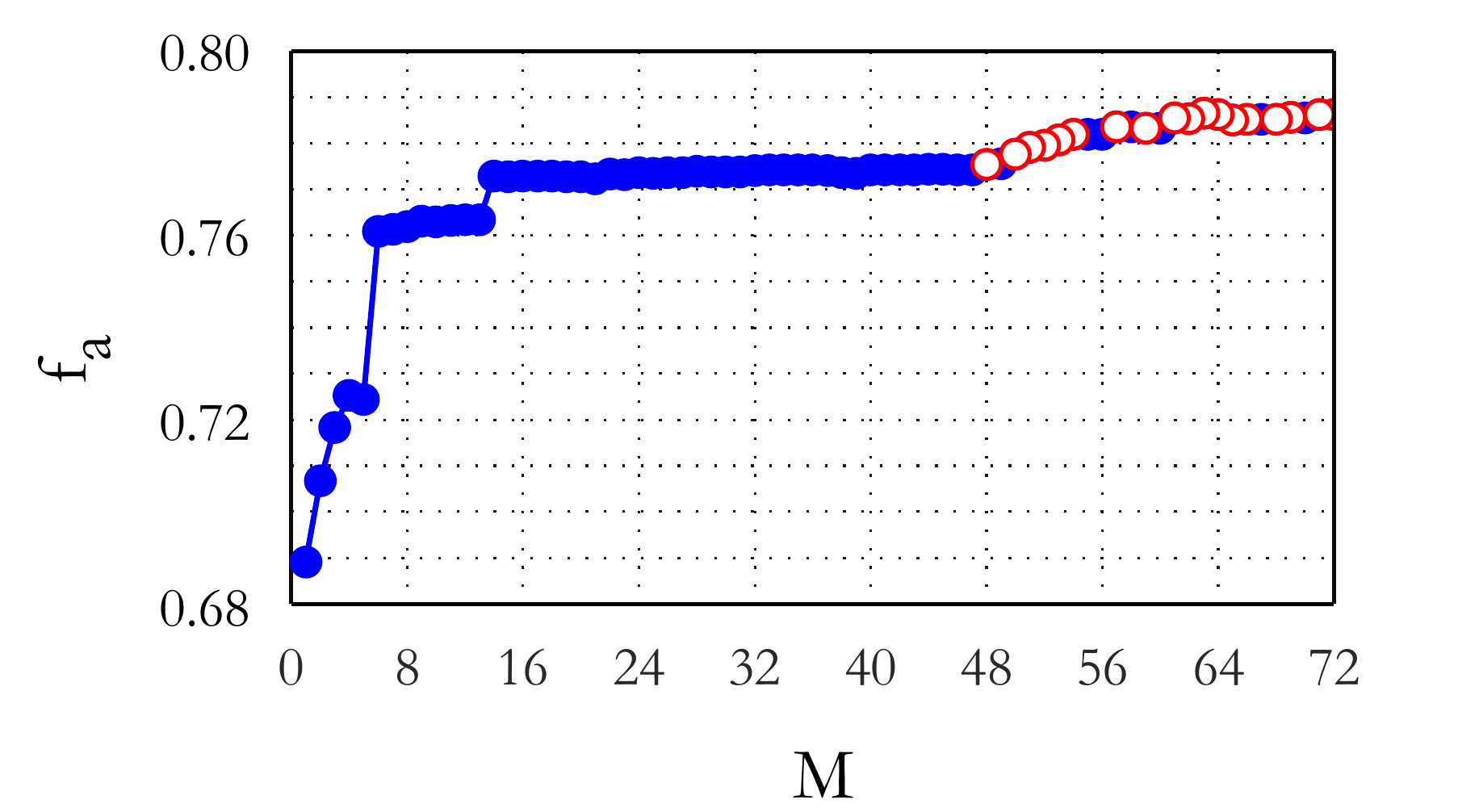}
		\caption[]{\label{fig5}
Recursive feature elimination determination of the most important structure functions. The fraction of training set atoms which are accurately classified, $f_a$, decreases as structure functions are eliminated. The color of the symbol shows whether the least important SF - which then gets eliminated - is a radial SF (red hollow circle) or angular SF (blue solid circle).
}		
	\end{center}
\end{figure}

\subsection*{Discussion}

In summary, we have used machine learning to introduce a structural quantity, softness, that is strongly correlated with the dynamics of atomic rearrangements in grain boundaries. 
Correlations between local structure and rearrangements are so strong in these nanocrystalline metals that over 96\% of the observed atomic rearrangements correspond to atoms for which $S>0$. 
The only information that enters the machine learning is provided by the training set, chosen to represent atomic environments  just about to rearrange and those that do not rearrange. A binary classification on this training set yields a remarkably rich lode of information. 
Softness can distinguish between certain GBs where rearrangements are imminent  from those which are markedly static and can reveal differences even within a given GB.
It also gives information about the energy landscape of the system, providing both the typical energy barriers and the attempt frequencies for atoms to rearrange - based solely on the their local structural environments. Such information is computationally expensive via existing methods.
Interestingly, we show (see SI Appendix) that while softness correlates to some degree with other structural parameters that identify defects in crystals, it is remarkably superior in identifying which defects/local environments are active in GB dynamics.

Our conclusion that the variation of entropy plays a central role in the atomic rearrangements at GBs is reminiscent of prior observations about nanoconfined viscous fluids.
Specifically, in Refs.~\cite{doi:10.1021/jp071369e,PhysRevLett.111.235901}, it was found that the total thermodynamic entropy of the confined fluid\footnote{Specifically the excess entropy, the entropy above that of an ideal gas at the same density and temperature.} closely reflects the rate of structural relaxation.
The atoms in GBs are somewhat analogous to a confined fluid, noting that the complicated confinement in GBs is due to structured surfaces of crystalline grains which can themselves rearrange and the configurations are not in equilibrium.
Our approach attempts to infer the local contributions to the entropy that are involved in the rearrangements.
The observation from Fig.~\ref{figSigmaEbarrier} that structures which tend to rearrange more frequently are associated with higher entropy constitutes a local version of the statement that the entropy underlies the relaxations.
However, the characteristic local energy barrier additionally emerges from our analysis of atomic dynamics.

Our approach shows that focusing on the dynamics at the atomic scale can serve as a viable alternative to the classification of environments based on the rich and complex zoology of crystallographically-allowed defects. This approach, successfully applied originally to glasses,~\cite{Schoenholz2016Structural,schoenholz2017relationship} 
suggests that it is possible to construct a single framework to describe atomic-scale dynamics in systems with varying degrees of order/disorder.
At the very least, this approach is complementary to existing methods that strive to relate dynamic materials phenomena to the underlying structure of their hosts. More optimistically, we note that the larger-scale dynamics are dictated by the interplay of softness with atomic rearrangements--while softness predicts the propensity to rearrange, a rearrangement alters local structure and hence softness. Understanding this interplay is the first step towards constructing a theory of plasticity that has the potential to span the entire gamut of materials from crystalline to glassy.

\nocite{Thomas2016}
\nocite{Lazar2011}
\nocite{Lazar2015}
\nocite{PlimptonFast1995}
\nocite{Chang2011SVM}
\nocite{cubuk2015identifying,CubukSchoenholz2016,Schoenholz2016Structural}
\nocite{BehlerParinello2007}
\nocite{Schoenholz2016Structural}
\nocite{Keys2011Excitations,SmessaertRottler2013}
\nocite{Schoenholz2016Structural,CubukSchoenholz2016}
\nocite{CubukSchoenholz2016}
\nocite{Stukowski2010Ovito,PlimptonFast1995}
\nocite{CubukSchoenholz2016}

\noindent\rule{8.5cm}{0.4pt}

We thank the UPenn MRSEC, NSF-DMR-1720530 (TAS), as well as  computational support provided by the LRSM HPC cluster at the University of Pennsylvania (TAS), the DOE GAANN program, grant number P200A160282 (SLT) and XSEDE via NSF grant number ACI-1053575 (SLT) \cite{xsede}, the US DOE, Office of Basic Energy Sciences, Division of Materials Sciences and Engineering under Award DE-FG02-05ER46199 (SSS,AJL), the US NSF Division of Materials Research DMR-1507013 (DJS) and the Simons Foundation (327939 to AJL).


\end{document}



\title{Supplemental Information}

%

\maketitle

	\section{Molecular Dynamics Simulation of Poly-crystalline Aluminum}
	\label{SupplementalMD}
	
	For completeness, we give additional details of the molecular dynamics simulations described in the main text.  The simulations follow a procedure developed in Ref.~\citenum{Thomas2016}.
	Simulations of nanocrystalline aluminum are initialized using a Voronoi tesselation generated from a Poisson distribution of points, which then evolves according to a curvature-flow grain growth simulation method~\cite{Lazar2011}. This method is chosen because Voronoi-Poisson microstructures exhibit grain size distributions and grain shapes (e.g., flat GBs with unrealistic GB junction angles) that differ greatly from those seen in grain growth~\cite{Lazar2015}. After the microstructure is created in this manner, each grain is populated with atoms in a face-centered cubic (FCC) lattice where the orientation of the lattice in each grain was chosen at random.
	In each simulation there are roughly 4 million atoms constituting approximately 550 grains with 6000 grain boundaries.
	Simulations proceed at fixed temperature with a Nos\'e-Hoover thermostat\cite{PlimptonFast1995} for at least 130 ps, then continue for an additional 10 ps within a zero pressure NPT ensemble during which the atomic trajectories are stored for analysis.

	\section{Support Vector Machine}
	
	We use a support vector machine~\cite{Chang2011SVM} (SVM) to reduce the high-dimensionality of atomic configurations into a binary classification of the atoms' likelihood to rearrange. We then introduce a continuous, signed, scalar quantity, softness, based on the binary classification.
	The method was developed in previous work \cite{cubuk2015identifying,CubukSchoenholz2016,Schoenholz2016Structural} for application to glassy systems.
	
	First, the positions of all the atoms surrounding a given atom $i$ are reduced to a set of $M$ structure function values, as discussed in the main text.
	We use the structure functions~\cite{BehlerParinello2007}  
	\begin{equation*}
	R_{i\alpha} = \sum_j \text{exp}(-(d_{ij} - \mu_{\alpha})^2/(2\sigma_0^2))
	\text{,\hspace{5mm}}
	A_{i\beta} = \sum_{j,k} 
	(1+\lambda_{\beta} cos \theta_{ijk})^{\zeta_\beta}
	f_{ij}f_{jk}f_{ki}
	\text{,\hspace{5mm}}
	f_{ij} =
	\text{exp}(-\eta_{\beta} d^2_{ij})
	\text{cos}(\pi d^2_{ij}/d_0+1)
	\text{.}
	\end{equation*}
	A radial structure function $R_{i\alpha}$ encodes information about the number of neighbors at different distances from the atom $i$.  
	An angular structure function $A_{i\beta}$ encodes the local structure using functions of the angles and perimeters of triangles that can be formed with atom $i$ and pairs of its neighboring atoms.
	$d_{ij}$ is the distance between atoms $i$ and $j$, and $\theta_{ijk}$ is the angle formed with the vector from atom $i$ to $k$.
	The sums extend over neighboring atoms of $i$ out to a cutoff distance of $d_0$.
	Parameter values vary with $\alpha$ and $\beta$ and are provided in Table~\ref{tab:params};
	we use $M = 72$ structure functions, consisting of 18 radial structure functions ($\alpha = \{ 1,2, \ldots 18 \}$) and 54 angular structure functions ($\beta = \{ 1,2, \ldots 54 \}$).

	\begin{table}[h]
		\centering
		\begin{tabular}{|l|l|}
			\hline
			$\alpha$ & $ 1,2,3, \ldots 12,13,14, \ldots 17, 18  $ \\ 
			\hline
			$\mu_\alpha$ (\AA) & $ 2.6, 2.7, 2.8, \ldots 3.7,3.8,4.0, \ldots 4.6, 4.8 $ \\ 
			\hline \hline
			$\beta$ & $ 1, 2, \ldots 27, 28, 29, \ldots 54 $ \\ 
			\hline 
			$\eta_\beta$ ($\text{\AA}^{-2}$) & $0.61, 0.49, \ldots \text{e}^{-\sqrt{\beta}/2}, \ldots 0.025$  \\			
			$\zeta_\beta$ & $ 0.1, 0.2, \ldots 2.7, 0.1, 0.2, \ldots 2.7 $ \\
			$\lambda_\beta$ & $ 1,1,\ldots 1,-1,-1, \ldots -1 $  \\	
			\hline
		\end{tabular}
		\caption{\label{tab:params}
			Parameter values used in the structure functions.
			$\mu_\alpha$ varies approximately from the nearest neighbor spacing, $d=2.8$~\AA,  to the cutoff distance, $d_0=5$~\AA, and $\sigma_0=0.25~\text{ \AA }$.
			$\eta_\beta$ and $\zeta_\beta$ span a range near 1.0 and are included twice, once with $\lambda_\beta=1$ and once with $\lambda_\beta=-1$.
		}
	\end{table}
	
	SVM is a supervised machine learning method.  This means that data in the training set are labeled; in our case, the atomic structures are labeled as belonging to one of two groups based on whether they will rearrange.  We choose these atomic structures from simulations conducted at $T=T_m/2$. The first subset consists of atomic structures that are about to undergo a rearrangement, while the second subset consists of atomic structures that are in a long period without rearrangements.

	\begin{figure}
		\begin{center}
			\includegraphics[width=.4\linewidth]
			{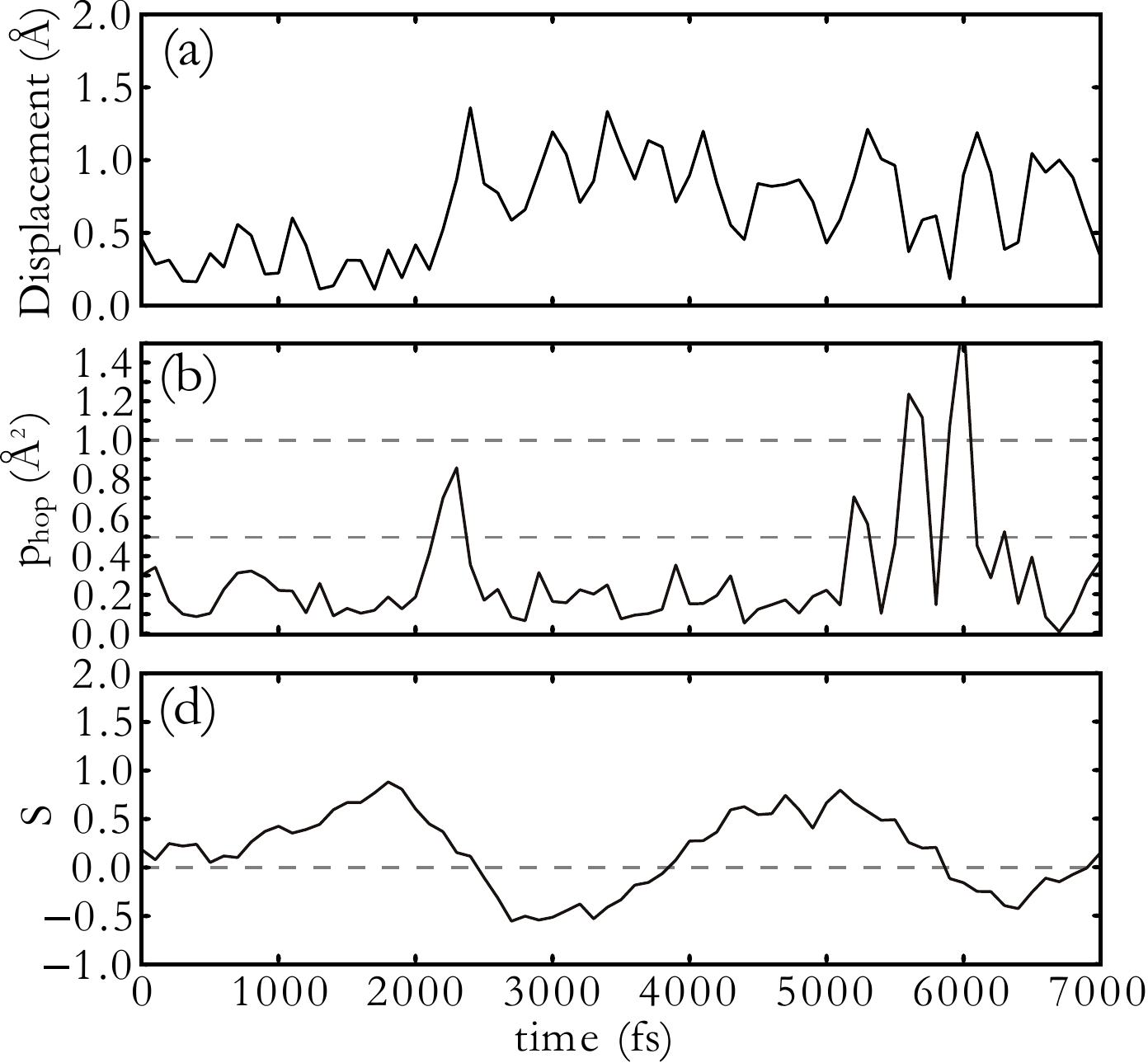}
						\includegraphics[width=.380\linewidth]
			{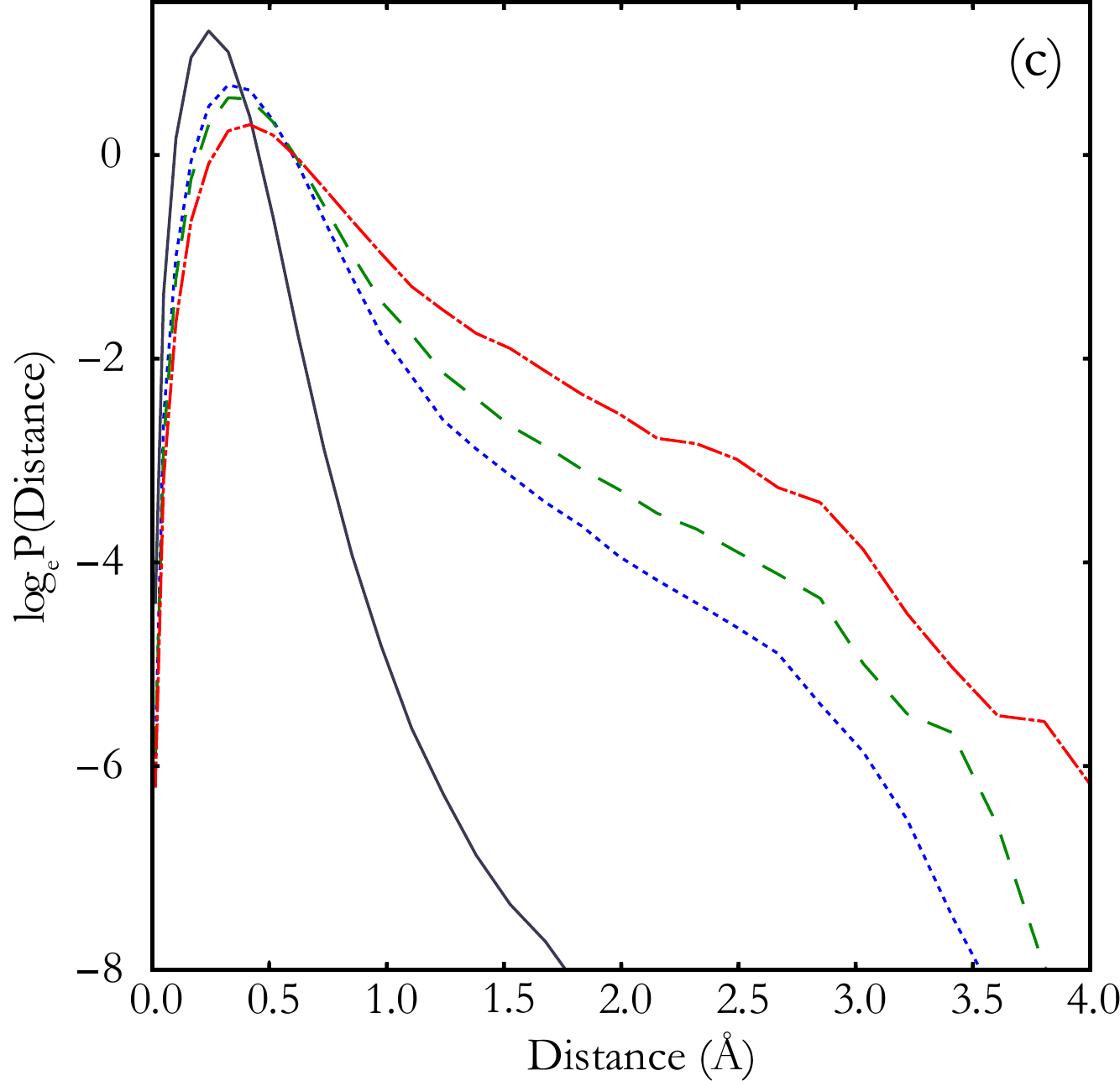}
			\caption[]{
				\label{atomictraj}
				(a-c) A sample trajectory illustrates the indicator function $p_{hop}$ and the quantity softness. (a) The displacement of an atom near a grain boundary remains low for several picoseconds before hopping to a new location.  (b) This is accentuated in the plot of $p_{hop}$.  Analysis of the mean-squared-displacement of trajectories such as these motivates the choice of the $p_{hop}$ parameters (the high and low thresholds, $p_c$ and $p_c/2$, indicated by dashed lines, and the window time $t_R = 400$ fsec) though results are insensitive to these cutoffs.  (c) The distribution of distances diffused by atoms during a full simulation (10 psec) shows that $p_{hop}$ discriminates between those atoms that can move a significant distance and those that cannot.
				Atoms that are never detected to hop, \emph{i.e.} atoms for which $p_{hop}$ never exceeds $p_c/2$, (solid back) generally remain localized; 87\% of these atoms end up within 0.4$\text{\AA}$ of their initial position. Atoms that hop one time (dotted blue) ($p_{hop}$ exceeds $p_c/2$ exactly one time during the simulation), two times (dashed green), or four times (dashed-dotted red) diffuse greater distances.
				Note that a hop does not necessarily result in the atom moving a significant distance in the end, because atoms can commonly rearrange to a metastable configuration and rearrange back within the time $t_R$ if nearby atoms do not also happen to rearrange. The percentage of atoms that remain within 0.4 $\text{\AA}$  of their initial position is 54\% for atoms that hop once, 48\% for atoms that hop twice, and 36\% that hop three times.
				(d) The softness of the atom, (with 900-fsec boxcar smoothing to reduce fluctuations) increases before the hop. 
		}
		\end{center}
	\end{figure}

	To identify atoms for these two training subsets, we follow previous work~\cite{Schoenholz2016Structural} using an indicator function, $p_{hop}(t')$~\cite{Keys2011Excitations,SmessaertRottler2013}:
	\begin{equation}
	p_{hop}(t')=\sqrt{ \langle ( \vec{r} - \langle \vec{r} \rangle_B )^2 \rangle_A
		\langle ( \vec{r} - \langle \vec{r} \rangle_A )^2 \rangle_B } .
	\end{equation}
	Here, the trajectory $\vec{r}(t)$ of an atom is averaged over the first half (A) of a moving time window, producing $\langle \vec{r} \rangle_A$.
	The moving time window has duration $t_R$ and is centered at $t'$.  
	We choose $t_R/2$ to be 200 fs based on 
	an analysis of typical atomic trajectories
	showing a crossover from ballistic to caged motion on this timescale.
	
	To calculate $p_{hop}$, we evaluate $\langle ( \vec{r} - \langle \vec{r} \rangle_A )^2 \rangle_B$, which is the mean squared distance from the position during the second half (B) of the time window to the position $\langle \vec{r} \rangle_A$. $p_{hop}$ is the time-symmetrized version of this quantity. Fig.~\ref{atomictraj} illustrates that $p_{hop}$ identifies rearrangements. An atomic trajectory is shown in Fig.~\ref{atomictraj}(a) and the corresponding $p_{hop}$ value is shown in Fig.~\ref{atomictraj}(b).
	$p_{hop}$ shows that the atom has undergone a large change of displacement, indicative of a rearrangement.
	We consider a clear rearrangement to occur if $p_{hop}$ exceeds a threshold $p_c=1.0 \text{\AA}^2$.  Since there is a broad spectrum of atomic motions, the precise values used are somewhat arbitrary, and we have confirmed that the results are insensitive to a 50\% increase or decrease in this threshold. Fig.~\ref{atomictraj}(c) shows the distribution of distances diffused by atoms during the simulation. Atoms with zero detected rearrangements generally do not diffuse significant distances. Atoms with one detected rearrangement sometimes end up back at their initial position; this is usually due to a first hop followed by a second hop back within 200 fsec and so is only identified as a single rearrangement.  However, if other nearby atoms themselves rearrange and impede a hop back, or the new rearranged configuration is sufficiently stable, then the atom remains at the new location.  Additional rearrangements may propagate the atoms diffusively to greater distances.
	
	Specifically, an atom at time $t$ is eligible for the first group of the training set if its  $p_{hop}$ exceeds the threshold $p_c$ at $t+t_R/2$.  The second group in the training set was selected by first identifying which atoms do not exceed a low threshold $p_q= p_c/2$ over a long time interval $\ge t_q$.
	These long quiescent intervals are exhibited by most atoms in grain interiors, away from defects, as well as many atoms in the grain boundary.  Atoms in the middle of a quiescent interval are eligible for the second group of the training set. 
	From plotting of trajectories like that shown in Fig.~\ref{atomictraj}, we see that $p_{hop}$ usually remains less than 0.5 $\text{\AA}^2$ for long intervals.  We therefore choose $p_q=0.5 \text{\AA}^2$.  We use $t_R = 400$~fs and $t_q = 1.8$~ps, but find that our results are insensitive to these choices.

	We choose $N= 10,000$ atoms from among the eligible atoms for each subset. This represents a small fraction of the eligible atoms. Those in the first subset are assigned a label $y_i=1$ while those in the second subset are assigned $y_i=-1$. To train the SVM, we first introduce an orthogonal axis for each of the $M$ structure functions so that the local environment of each atom $i$ is represented as a point in this $M$-dimensional space. 
	The values from each structure function are shifted to a mean of zero and scaled to have unit variance.
	We then construct the hyperplane $\vec{w}\cdot \vec{x} - b = 0$ that best separates $y_i=1$ atoms from $y_i=-1$ atoms. This is the hyperplane that minimizes  $\sum_{i=1}^{2N} \text{max} \left(0, 1- y_i \cdot (\vec{w} \cdot \vec{x}_i - b) \right)$ - that is, it minimizes the penalty of having atoms on the incorrect side of the hyperplane. This is a generalization of linear regression.			
	
	Thereafter, the SVM may be used as a binary classifier; we refer to the side of the hyperplane that contains predominantly atoms with $y_i=1$ as the ``soft" side, and to the other side of the hyperplane, with atoms that predominantly have $y_i=-1$, as the ``not soft" side. To check the prediction accuracy of the hyperplane, we conducted cross-validation with 10 folds. In other words, we divided our training set into 10 equal subsets, each containing equal numbers of atoms with $y_i=\pm 1$. We then constructed a hyperplane from 9 of the 10 subsets and computed the fraction of $y_i=1$ atoms in the remaining subset that lie on the soft side of the hyperplane, and the fraction of $y_i=-1$ atoms that lie on the not-soft side. Finally, we averaged these fractions over the all of the 10 possible choices of the 9 subsets chosen for constructing the hyperplane. We find that 96\% of atoms with $y_i=1$ lie on the soft side of the hyperplane, while 90\% of atoms with $y_i=-1$ lie on the non-soft side of the hyperplane. Thus, the hyperplane  distinguishes effectively between $y_i=\pm 1$.
	
\begin{figure}
		\begin{center}
			\includegraphics[width=.3\linewidth]
			{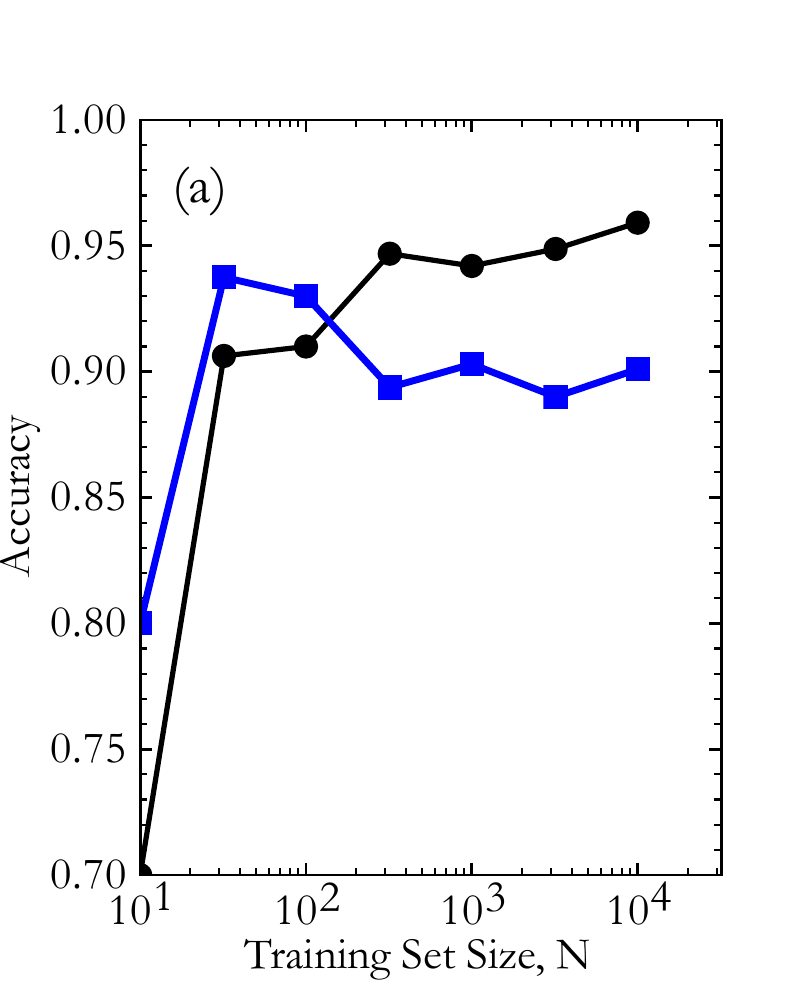}
			\includegraphics[width=.3\linewidth]
			{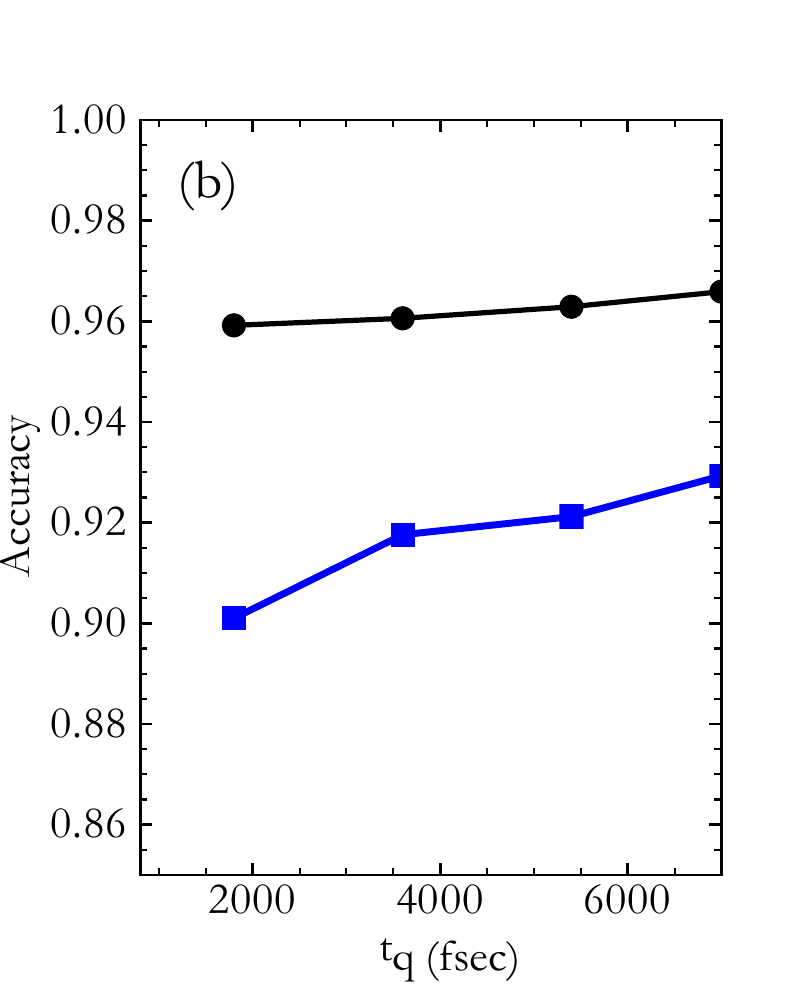}
			\includegraphics[width=.3\linewidth]
			{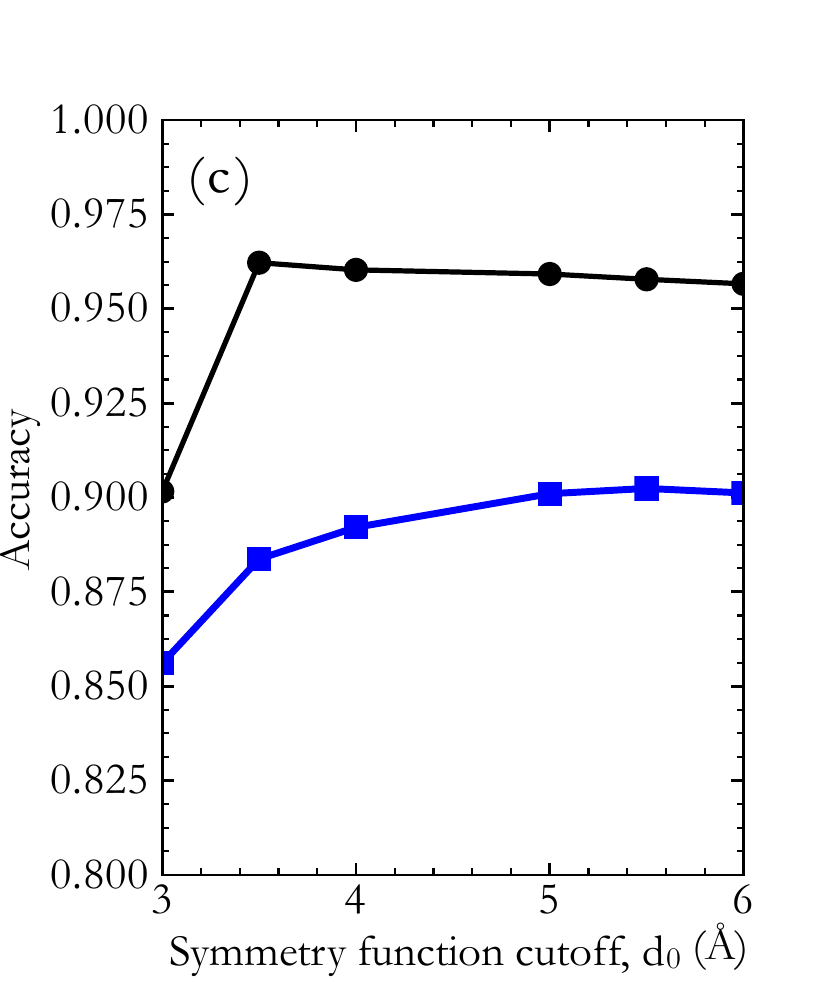}        
			\caption[]{
				\label{CVaccuracy}
				Cross validation with 10 folds shows greater than 90\% accuracy when the SVM is used as a binary classifier.   Prediction accuracy is slightly higher for atoms with $y_i=1$ (black circles) than for atoms with $y_i=-1$  (blue squares).  Plots vary training set size $N$, quiescent time $t_q$, and distance cutoff of the structure functions $d_0$.  When not varied, the values are $N=10,000$, $t_q=1800$, and $d_0 = 5$ \AA.
			}
		\end{center}
	\end{figure}

	Fig.~\ref{CVaccuracy} shows how the cross-validation success rate varies with the SVM and structure function parameters.
	The overall cross-validation accuracy becomes somewhat insensitive to the training set size already by $N=100$, with a steady slight increase for $y_i=1$  atoms. 
	We also tested training on only specific subsets of atoms of the polycrystal, though this had only a trivial effect on the results.
	Specifically, training only on grain boundary atoms (CNA-crystalline atoms removed) resulted in a hyperplane nearly parallel to the original.  This shows that
	the training set includes enough atomic configurations, and focusing training on atoms in the grain boundary alone does not improve the identified hyperplane. The position of the hyperplane shifts along the normal direction when only grain boundary atoms are included  because the training set  no longer includes atoms in FCC grains that never rearrange.

	Once we have the hyperplane, we calculate the structure functions for each atom in our simulations and locate the position of each atom's local structural environment as a point in the  $M$-dimensional space of structure functions. 
	We define the distance from the point corresponding to atom $i$ to the hyperplane as the magnitude of the``softness" $S_i$ of atom $i$. The sign of $S_i$ tells us which side of the hyperplane the point lies on; if the point lies on the soft side then $S_i>0$, and if it lies on the not-soft side then $S_i<0$.
	
	Fig.~\ref{atomictraj}(d) shows the softness of the example atom as a function of time during the atom's trajectory.  The softness is larger shortly before a hop occurs. After a hop the atom is in a more stable environment and the softness becomes negative.
	The local atomic environment slowly becomes soft again over the course of several picoseconds.
	The evolution of softness with time was studied statistically in the context of bulk 3D glasses\cite{Schoenholz2016Structural,CubukSchoenholz2016} in previous work.

\section{Determination of most important structure functions}

We investigate which structure functions (SFs) are connected to likelihood to rearrange. In this section the training sets ($N=10,000$) are composed only of non-crystalline (CNA) atoms to focus solely on grain boundary atoms.

\begin{figure}
	\begin{center}
		\includegraphics[width=0.8\linewidth]{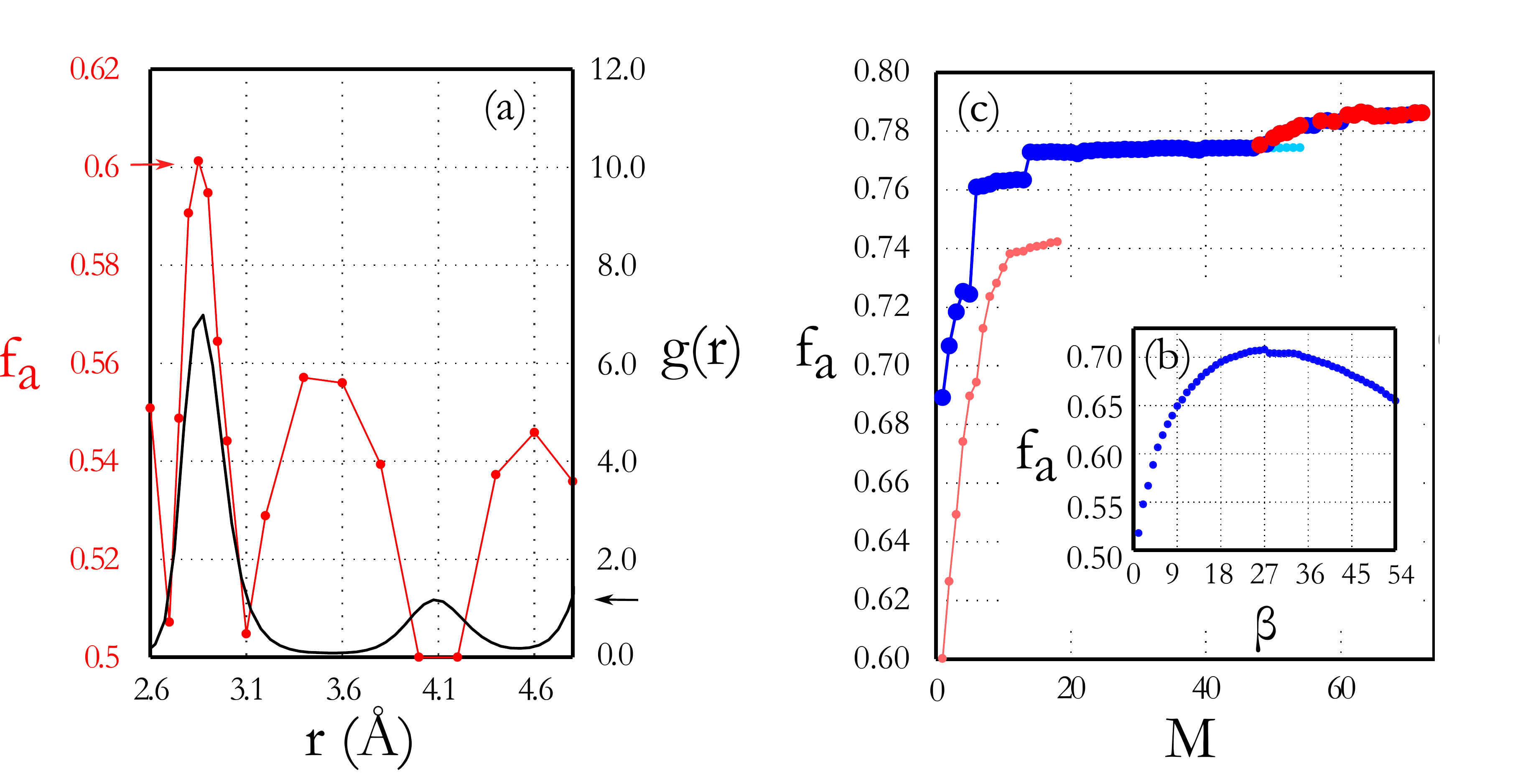}
		\caption[]{\label{figS3}
The contribution of specific structural features to softness can be seen by how strongly each structural descriptor function (SF) discriminates particles that will rearrange from those that will not.
(a) Plotted is the accuracy $f_a$ (small red circles) with which the SVM, trained with a single radial SF centered at $\mu_\alpha=r$, identifies which GB atoms will rearrange. The radial distribution function, $g(r)$ of the GB atoms is also plotted.
(b) Plotted is the accuracy (small blue circles) from training with a single angular SF of specified ID.
(c) The decrease in accuracy during RFE is shown. The color of the symbol shows whether the least important SF - which then gets eliminated - is a radial SF (red) or angular SF (blue). Large circles show RFE using all SFs of both types, while small circles show RFE begun exclusively with only the radial or angular SFs.
		}		
	\end{center}
\end{figure}

We train the SVM using a single SF to find the fraction of the training set labels ($y_i$) that can be correctly identified on the basis of that SF alone. 
We first consider radial SFs, $R_{i\alpha}$, defined above, which are approximately proportional to the number of atoms at radial distance $\mu_\alpha$. Figure~\ref{figS3} shows $f_a$, the fraction of training set atoms that are accurately classified, from considering only one radial SF, centered at distance $\mu_\alpha=r$. 
The number of neighbors at the first peak of the radial distribution function, $g(r)$, is most informative, allowing 60\% accuracy. The sign of the hyper-plane weight is negative, meaning that having fewer neighbors at that distance will increase softness. We note that this is consistent with other evidence mentioned in the main text that low-softness atoms have more crystalline neighbors.

The number of atomic neighbors at a \emph{valley} of $g(r)$ is also informative, allowing prediction with 56\% accuracy. The sign of the hyper-plane weight is positive here; softness increases with the number of atoms. The presence of atoms at a valley of $g(r)$ indicates an atypical environment and increased likelihood to rearrange as was observed previously in bulk glasses\cite{Schoenholz2016Structural}.

We next consider SFs $A_{i\beta}$ that depend on the angles formed with pairs of neighbors around the central atom.
While less easily interpretable, these SFs encode angular information beyond the radial distribution of neighbors.
These SFs were found to be relatively unimportant in bulk glasses\cite{Schoenholz2016Structural}.
In GBs, in contrast, Fig. 5(b) and (c) show that almost all angular SFs are much more predictive than radial SFs.
Fig. 5(b) shows that a single angular SF can yield up to 71\% accuracy. Almost any angular SF (50 of the 54 used) produces accuracy $> 60\%$, and is therefore more linked to softness than any radial SF.

The importance of the angular SFs is confirmed also in Fig~5(c) using recursive feature elimination (RFE).
RFE uses all SFs and iteratively eliminates the SF corresponding to the component of the weight $\vec{w}$ with smallest magnitude, targeting a smaller fingerprint of the highest accuracy.
As the number $M$ of SFs decreases, all radial SFs are eliminated rapidly (red circles), decreasing accuracy only from 79\% to 77\%.
Only angular SFs (blue circles) remain as the final $M=47$ SFs. 
For completeness, Fig~5(c) also shows $f_a$ from training with only radial SFs and only angular SFs.  Using many radial SFs (light red) increases $f_a$, showing that many radial SFs together start to capture similar information, but $f_a$ using angular SFs (light blue) remains greater.
Since angular SFs were relatively unimportant in studies on bulk glasses\cite{Schoenholz2016Structural}, we conclude that the explicit angular information contained in angular structure functions has greater utility for predicting rearrangements in GBs than in bulk glasses.

	\section{Correlations of softness with other local quantities}
	
\begin{figure}
		\begin{center}
			\includegraphics[width=.75\linewidth]
			{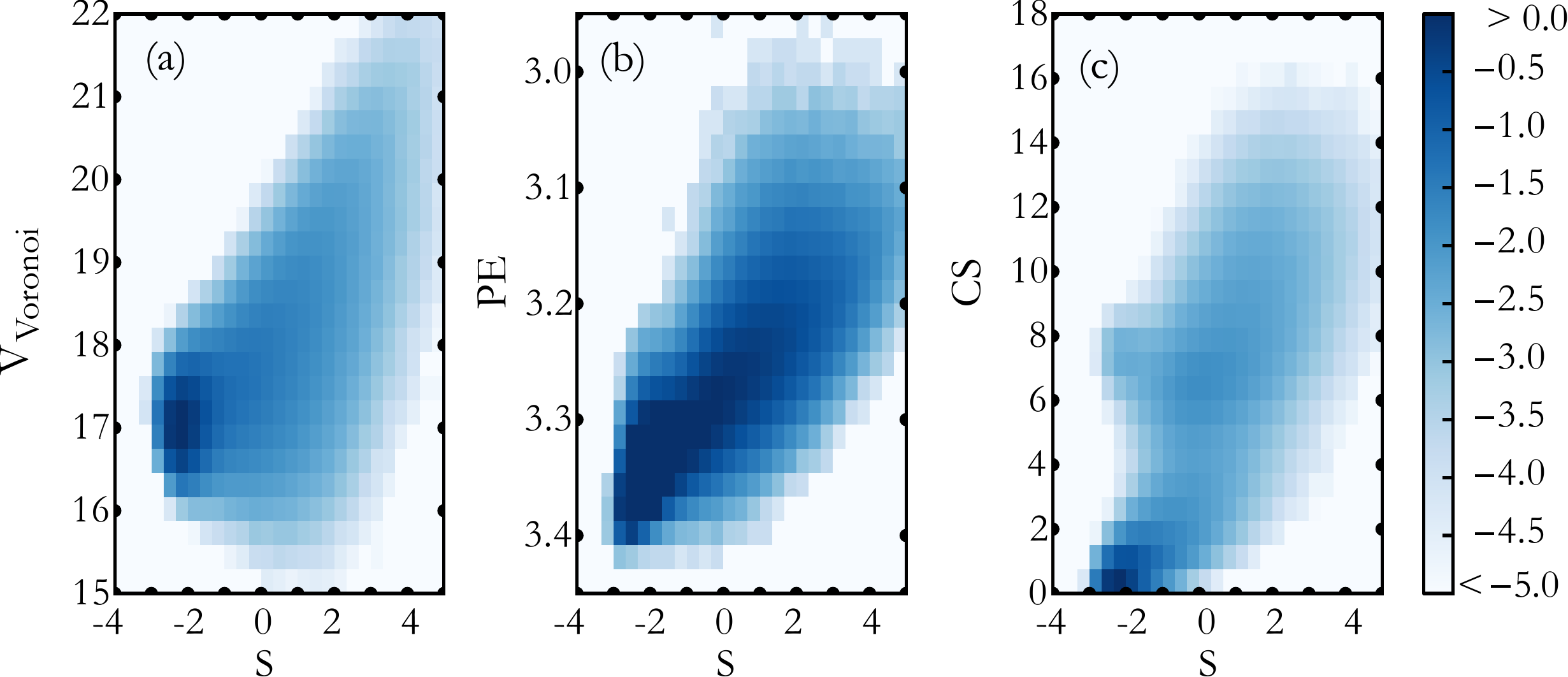}
			\caption[]{
				\label{correlations}
				Shown is the distribution of atoms with a particular softness and (a) Voronoi volume, (b) potential energy and (c) centrosymmetry. Soft atoms are generally in less symmetric environments, with higher energy, with a Voronoi volume that differs from that of the FCC crystal. The color denotes the value of the log of the normalized density function.
			}
		\end{center}
	\end{figure}

	Many local quantities have been used to describe the environment surrounding an atom. Fig.~\ref{correlations} shows the correlations between softness $S_i$ of an atom $i$ and some of the most common local quantities used, such as the volume of the Voronoi cell for atom $i$, its potential energy, and its centrosymmetry (defined below).    
	As observed before in 3D bulk glasses~\cite{CubukSchoenholz2016}, there are strong correlations between softness and other local quantities that have been found to correlate with likelihood to rearrange.
	In our system, Fig.~\ref{correlations}(a) shows that atoms of low softness have a Voronoi cell volume near the FCC value of 17 $\text{\AA}^3$.  Atoms with Voronoi volume greater than 19 $\text{\AA}^3$ are essentially all soft.  Likewise, the potential energy of atom $i$ is positively correlated with its softness, though there remains large spread for a given energy value (see Fig.~\ref{correlations}(b)).
	
	Centro-symmetry is a standard quantity used to characterize local structure around an atom \cite{Stukowski2010Ovito,PlimptonFast1995}.
	First, the $N_n=12$ atoms nearest to atom $i$ are identified.
	Next, all $N_n (N_n-1)/2$ possible pairings of those atoms are considered.
	For each pairing $(j,j^\prime)$, the squared magnitude of the vector sum of the positions relative to atom $i$ is computed: $U_{i,j,j^\prime}=|\vec{u}_{ij} + \vec{u}_{ij^\prime}|^2$.
	Note that $U_{i,j,j^\prime}$ is small when atoms $j$ and $j^\prime$ are nearly opposite the center atom.
	The centro-symmetry of atom $i$ is defined as
	\begin{equation*}
	\text{CS}_i = \sum_{j,j^\prime} U_{i,j,j^\prime}
	\end{equation*}
	where the sum occurs over the $N_n/2$ smallest $U_{i,j,j^\prime}$ values.
	Clearly, CS vanishes for every atom in the perfect FCC crystal, since 6 opposing pairs of atoms surround the center atom.
	
    	\begin{figure}
		\begin{center}
			\includegraphics[width=.4\linewidth]
			{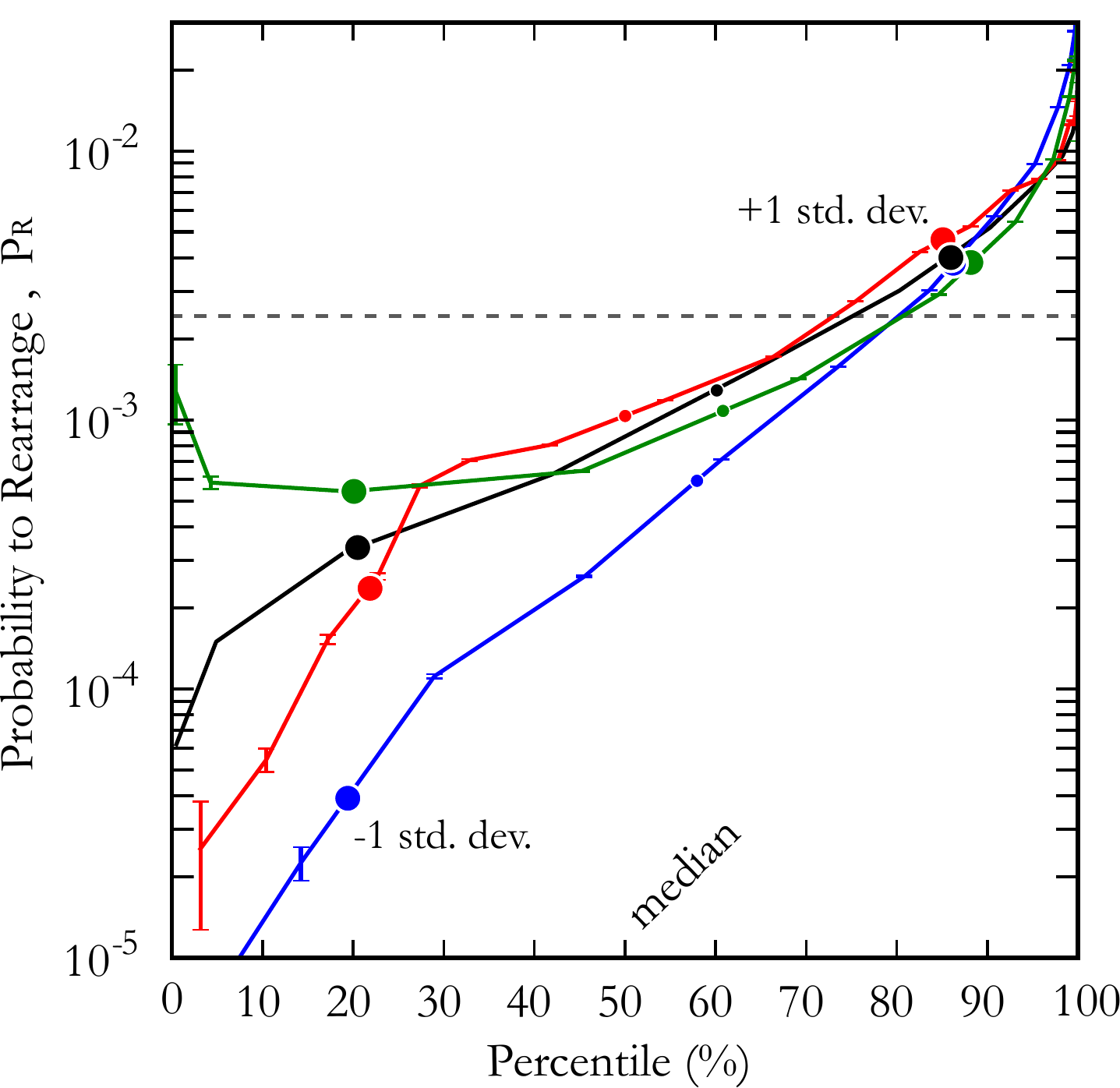}
			\caption[]{
				\label{generalQ}
				Softness is more predictive of likelihood to rearrange than other investigated quantities. For an atom in a grain boundary in the simulation, the probability that the atom will rearrange, $P_R$, is plotted \emph{vs.} the atom's percentile rank in terms of the value of its softness (blue). Similar curves are $P_R$ \emph{vs.} potential energy (black), centro-symmetry (red), and Voronoi volume deviation (green).
				For a quantity that is unrelated to rearrangements, atoms in each bin would be equally likely to rearrange (dashed line); quantities that are most predictive of rearrangements most strongly separate atoms by their likelihood to rearrange, \emph{i.e.} have the greatest slope.  The mean slope over the majority of particles is characterized by the value of $Q$ (Eq.~\ref{eq:Q}) which is the ratio of $P_R$ at $\pm 1$ standard deviation (large circles) from the mean (small circle).
			}
		\end{center}
	\end{figure}

	Fig.~\ref{correlations}(c) shows that the low-CS atoms atoms are also the least soft (recall that the CNA-identified FCC atoms are removed from the analysis).

	The utility of each quantity to predicting rearrangements is related to how strongly that quantity distinguishes atoms which often rearrange from those that rarely rearrange.
	If a quantity $X$ is unrelated to rearrangements, then the probability that a particular atom will rearrange is independent of its value of $X$.
	On the other hand, if the quantity $X$ is positively correlated with rearrangements, then $P_R$ will be somewhat higher for atoms with large values of $X$ than for small values of $X$.
	Following Ref~\cite{CubukSchoenholz2016}, for any candidate quantity $X$ that is continuous, we consider atoms one standard deviation ($\sigma_X$) above and below the mean value ($\mu_X$) and define the metric
	\begin{equation}
	\label{eq:Q}
	Q_X = \frac{P_R(X=\mu_X+\sigma_X)}{P_R(X=\mu_X-\sigma_X)}
	\end{equation}
	To focus on grain boundary dynamics, we filter out CNA-identified FCC atoms, so $\mu_X$ and $\sigma_X$ are the mean and standard deviation of $X$ values of atoms at the grain boundaries.
	Higher values of $Q_X$ indicate that $X$ is a better predictor of the propensity to rearrange. 
	We find that for this system, $Q=75$ for softness, $Q=6.5$ for the Voronoi volume deviation (\emph{i.e.} the magnitude of the difference of the volume from the average volume), $Q=18$ for local atomic potential energy, and $Q=53$ for centro-symmetry.
	This indicates that softness is considerably more predictive of rearrangements. Note also that the correlation of softness with CS is not noticeably stronger than for the Voronoi volume and potential energy. One cannot rely on the existence of a correlation with softness to quantify the ability of any given quantity to predict rearrangements.

	These results are not sensitive to the choice of evaluating $Q$ at one standard deviation above and below the mean.  This is apparent from Fig.~\ref{generalQ}, where $P_R$ is plotted as a function of each of the quantities.  So that different quantities may be compared, the horizontal axis of $X$ values has been mapped into the range $0$ to $100\%$ by sorting the atoms in the order of their value of X.
	Softness is steepest;
	compared to CS, softness produces a $P_R$ which is lower (by a factor of about 5.0 at the 20th-percentile) for low values and higher (by a factor of up to 2.1, above the 94th-percentile) at high values where most rearrangements occur.
	Circles indicate the location of one standard deviation above and below the mean value of $X$, so that the vertical separation between them on the logarithmic scale indicates the value of $Q$. 
	Clearly, softness most strongly separates atoms that are likely to rearrange at the high end from those that are unlikely to rearrange at the low end.

	\begin{figure}
		\centering
		\begin{minipage}{.45\linewidth}
			\centering
			\includegraphics[width=1\linewidth]
			{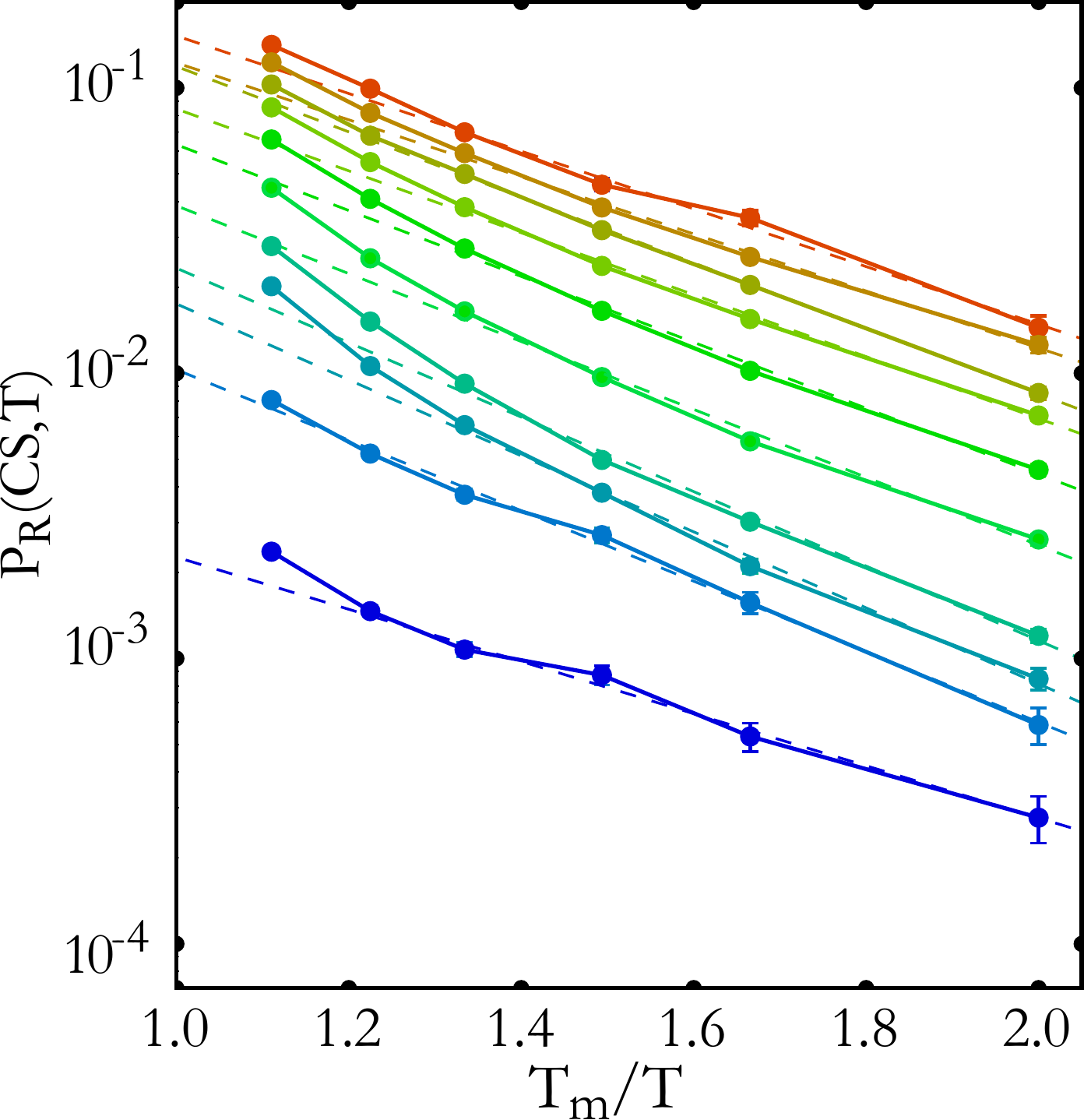}
			\caption[]{
				\label{CS_Arrhenius}
				The probability of rearrangement, $P_R$, \emph{vs.} inverse fraction of the melting temperature, $T_m/T$, for particular  centro-symmetry (CS) values.  CS values range from 3.2 (bottom curve) to 13.6 (top curve), with spacings of 0.8.
			}    
		\end{minipage}%
		\hspace{5mm}
		\begin{minipage}{.45\linewidth}
			\centering
			\includegraphics[width=1\linewidth]
			{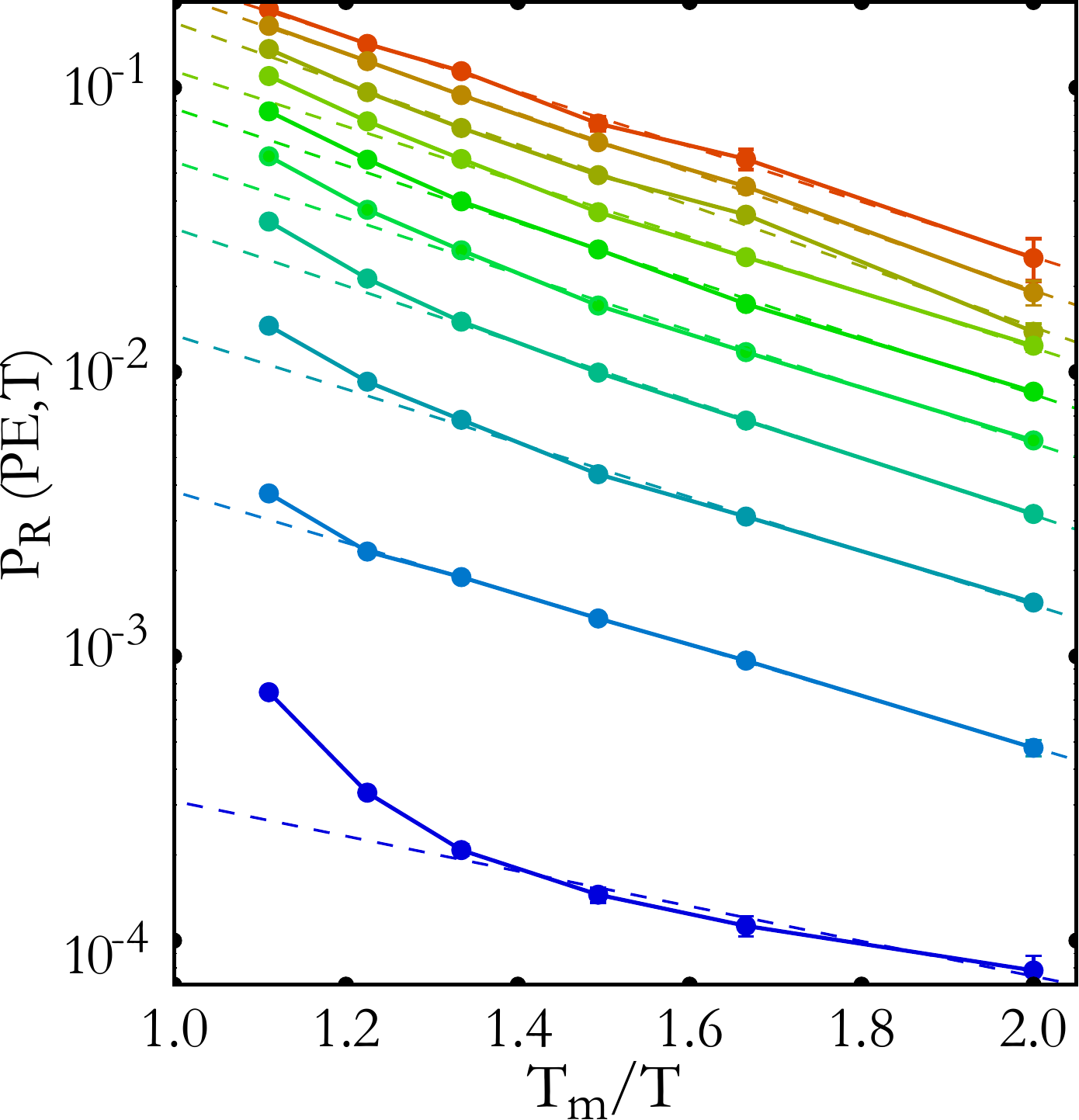}
			\caption[]{
				\label{PE_Arrhenius}
				The probability of rearrangement, $P_R$, \emph{vs.} inverse fraction of the melting temperature, $T_m/T$, for particular  potential energy values.  PE values range from -3.35 eV (bottom curve) to -3.08 eV (top curve), with spacings of 30 m eV.
			}
		\end{minipage}
	\end{figure}

	Although softness is more predictive of high- and low-likelihood to rearrange, $P_R$ rises significantly for all considered quantities in this system.
	For instance, the large value of $Q$ for centro-symmetry, and its correlation with softness, suggests that centro-symmetry may be used as a proxy for softness.
	
\begin{figure}
		\centering
		\begin{minipage}{.45\linewidth}
			\centering
			\includegraphics[width=.85\linewidth]
			{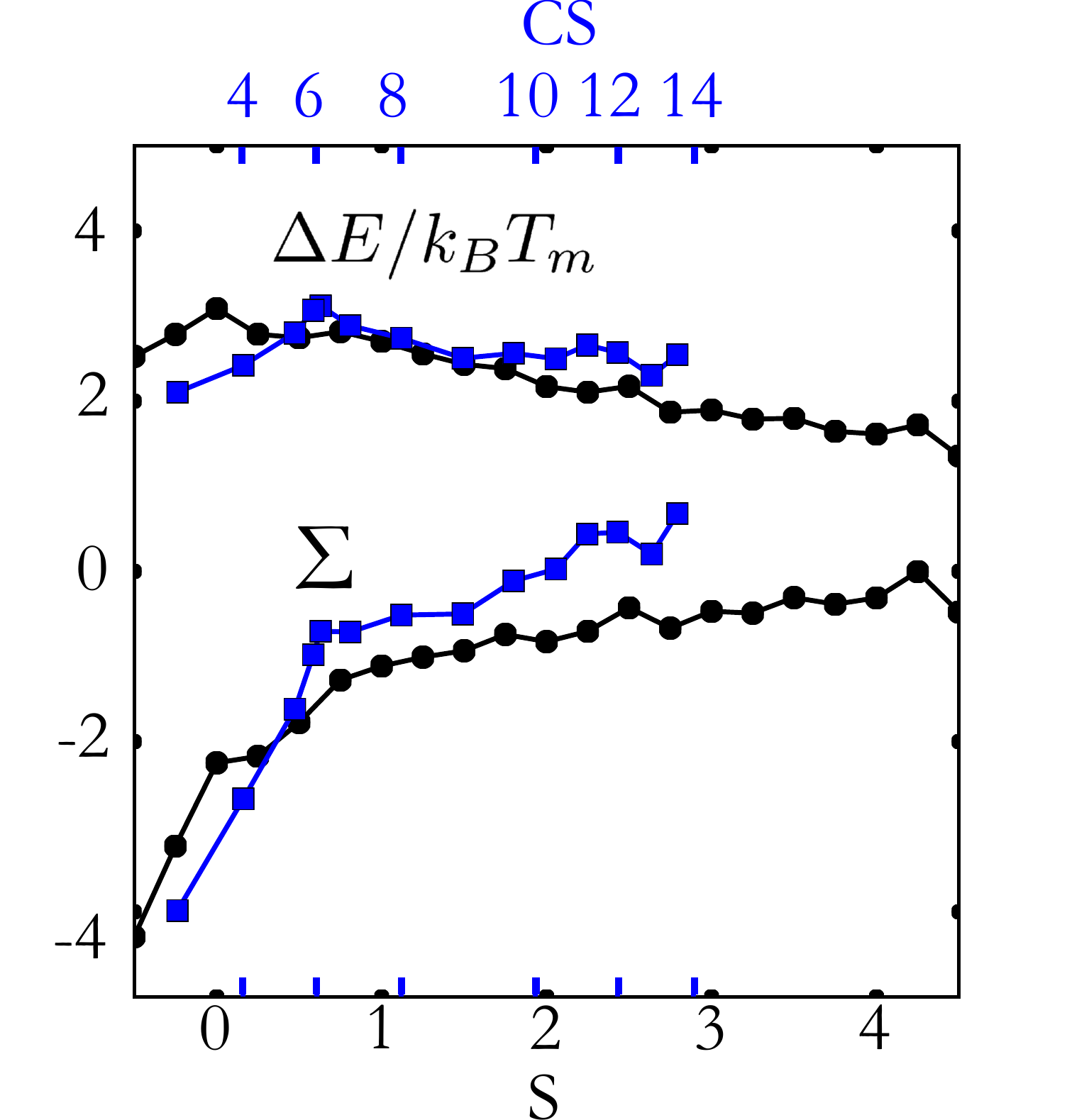}
			\caption[]{
				\label{fig:compareEbarrier}
				Energy barrier and $\Sigma$, as extracted from atoms of a particular softness (black circles) and extracted from atoms of a particular CS value (blue squares).  We simultaneously plot the average value of CS within a bin on the top axis and the average value of softness for the same atoms on the lower axis.
			}   
		\end{minipage}%
		\hspace{5mm}
		\begin{minipage}{.45\linewidth}
			\centering
			\includegraphics[width=.85\linewidth]
			{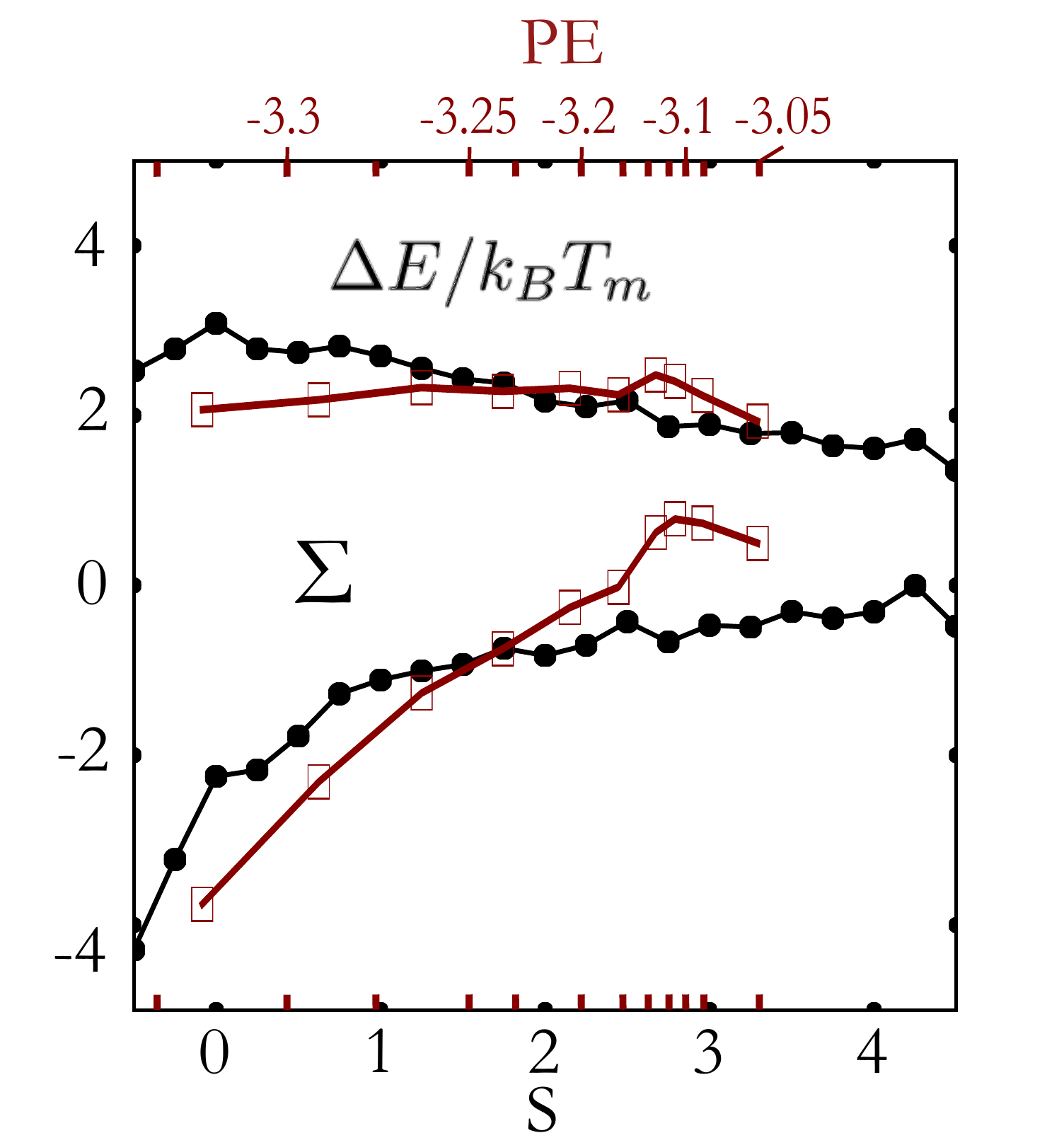}
			\caption[]{
				\label{fig:compareEbarrierPE}
				Energy barrier and $\Sigma$, extracted from atoms of particular softness (black circles) and extracted from atoms of a particular potential energy (blue squares). We simultaneously plot the average value of PE within a bin on the top axis and the average value of softness for the same atoms on the lower axis.
			}
		\end{minipage}
	\end{figure}

	Indeed, since $CS$ and potential energy are easier to calculate than softness, we have explored the possibility that the former quantities could be used to extract approximate information about the energy barriers.
	Fig.~\ref{CS_Arrhenius} shows that $P_R(\text{CS},T)$, the probability to rearrange for atoms of a given CS value, produces approximately Arrhenius behavior over a range of CS values.  Similarly, Fig.~\ref{PE_Arrhenius} shows that $P_R(\text{PE},T)$, the probability to rearrange for atoms of a given PE value, produces approximately Arrhenius behavior over a range of PE values. The greater linearity on these axes in Fig. 3 in the main text indicates the superior predictive accuracy of softness for extracting energy barriers.  However, if one insists on fitting the curves to Arrhenius behavior, the resulting best-fit $\Delta E$ and $\Sigma$ parameters (main text, Eq.~1) are reasonably similar for atoms binned by softness, CS, and PE, as shown in Figs.~\ref{fig:compareEbarrier} and ~\ref{fig:compareEbarrierPE}, giving greater confidence in the results for CS and PE.

	\section{Diffusion of atoms within a GB}

To quantify diffusion and softness in individual GBs, we first assigns IDs to all crystalline grains in the sample. For this purpose, we consider crystalline grain atoms to be those atoms that are further away than 3.0 $\text{\AA}$ of GBs (CNA-non-crystalline atoms). 
Clusters of these atoms 
are identified as individual grains. 
Next, we find atoms that are not part of any grain, but are within 9 $\text{\AA}$ of exactly two grains. 
These atoms are typically a quasi-planar collection of disordered atoms which may roughly be referred to as a single GB.
The number of atoms in a GB reach on the order of 10,000 atoms, and GBs smaller than 40 atoms are excluded from the analysis.

\begin{figure}
\begin{center}
	\includegraphics[width=.3\linewidth]
	{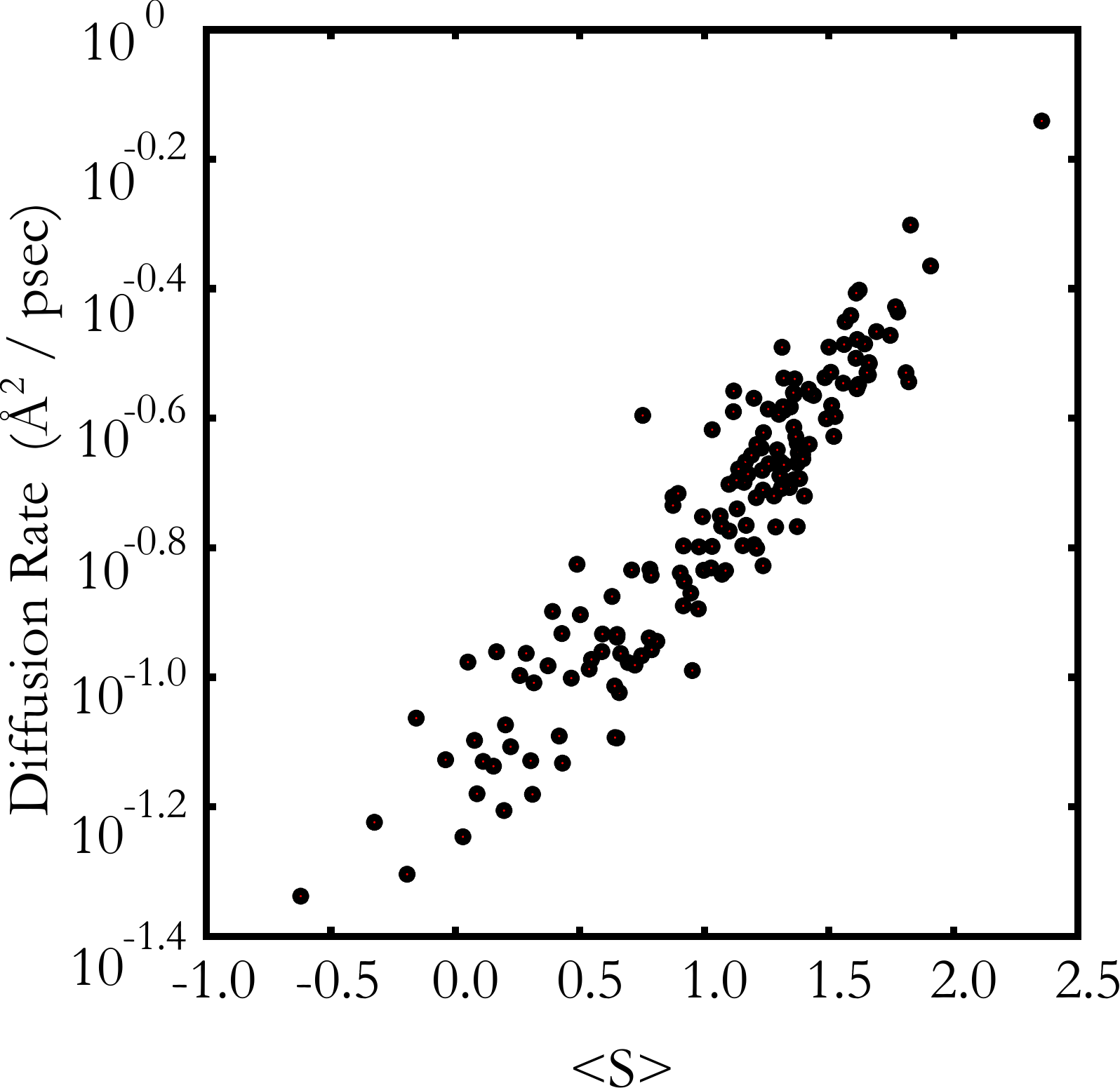}
	\caption[]{
		\label{figDvsS}
The diffusion of atoms within a GB increases with its mean softness.  The mean squared displacement (MSD) rate of GB atoms during a 10 psec simulation at $T=694K$ is plotted \emph{vs} the spatially-averaged  softness of the GB.
	}
\end{center}
\end{figure}

The diffusion rate of atoms within each GB is plotted against the mean softness $\langle S \rangle$ of each GB in Fig.~\ref{figDvsS}. 
Atomic diffusion increases approximately exponentially with softness of the GB. 
Note that the spatial structure of softness, which can indicate for example the presence of high-diffusivity paths within the GB, is not considered in this analysis of mean softness.
Nonetheless, the mean softness of the GB approximately determines the atomic diffusion.

